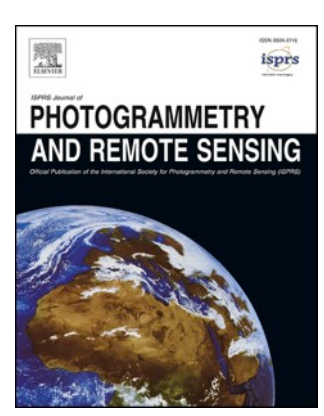

# Mapping melliferous tree species in Kenya via one-class classification with hyperspectral unsupervised domain adaptation

Zhaozhi Luo [a,c,*], Janne Heiskanen [a,b], Ilja Vuorinne [a,c], Ian Ocholla [a,c], Shiqi Zhang [d], Saana Järvinen [a], Xinyu Wang [e], Yanfei Zhong [f], Petri Pellikka [a,f]

[a] *Department of Geosciences and Geography, University of Helsinki, P.O. Box 64, 00014 Helsinki, Finland*
[b] *Finnish Meteorological Institute, P.O. Box 503, 00101 Helsinki, Finland*
[c] *Institute for Atmospheric and Earth System Research, University of Helsinki, P.O. Box 4, 00014 Helsinki, Finland*
[d] *School of Emergency Management, Xihua University, 610037 Chengdu, China*
[e] *School of Remote Sensing and Information Engineering, Wuhan University, 430079 Wuhan, China*
[f] *State Key Laboratory of Information Engineering in Surveying, Mapping and Remote Sensing, Wuhan University, 430072 Wuhan, China*

ABSTRACT

The beekeeping sector holds significant potential for livelihood diversification among the agropastoral communities in Kenya. Melliferous tree species play a critical role by providing essential nectar sources for bees. However, limited knowledge of their precise spatial distributions constrains the full development of beekeeping. One-class classification (OCC) offers a practical solution for detecting single target species without requiring extensive labeled data from other classes. Although existing OCC methods perform well in trained domains, the generalization capability to unseen domains remains limited due to domain shift. To address these challenges, this study proposes a hyperspectral unsupervised domain adaptation OCC framework (HyUDA-One) for tree species mapping using airborne hyperspectral imagery and laser scanning data. The spatial–spectral regularized pseudo-positive learning was designed to mitigate domain shift and improve model generalizability. The effectiveness of HyUDA-One was demonstrated by mapping three key melliferous tree species in two savanna landscapes in southern Kenya. The results show that HyUDA-One significantly improves performance in unlabeled domains. The F1-scores of 0.788, 0.845, and 0.768 were achieved for *Senegalia mellifera*, *Vachellia tortilis*, and *Commiphora africana* in the trained domain, respectively. In the untrained domain, the F1-scores of *Senegalia mellifera* and *Vachellia tortilis* were 0.756 and 0.884, respectively. The distribution maps revealed the spatial patterns of these melliferous tree species and the nectar source availability, offering an important reference for sustainable beekeeping development in savanna landscapes. Furthermore, the proposed framework can potentially be extended to other mapping applications, such as invasive species detection.

## 1. Introduction

Beekeeping has a long-standing tradition in Kenya, with various communities practicing traditional methods for centuries. Traditional practices typically involve suspending log hives to trees in natural woodlands. However, honey from harvested log hives is often unrefined and the productivity per hive remains low. Modern beekeeping was introduced by the colonial government in the 1950 s and has since been promoted as a strategy of livelihood diversification, particularly in the arid and semi-arid (ASA) regions (Affognon et al. 2015). Despite these efforts, beekeeping remains a relatively minor economic activity in Kenya, with the majority of the approximately 1.4 million beehives still being traditional log hives, which account for 80 % of the honey production (Siminyu et al. 2024). The sector operates far below its potential, producing only a fraction of the estimated 100,000 tons of honey annually (Carroll and Kinsella 2013). In addition to limited access to modern beekeeping knowledge and equipment, a major constraint is the lack of understanding on the spatial distribution of key nectar-producing tree species. Such understanding is essential for enhancing honey production and promoting the sustainable development of beekeeping in ASA landscapes. Several indigenous tree species, including *Senegalia mellifera*, *Vachellia tortilis*, and *Commiphora africana*, are vital to local apiculture, serving as key nectar sources for bees. Their spatial distribution and density directly influence apiary productivity and honey quality.

* Corresponding author.
*E-mail address:* zhaozhi.luo@helsinki.fi (Z. Luo).

Historically, tree species distribution mapping relied heavily on ground-based field surveys (Thijs et al. 2015), which are labor-intensive, time-consuming, and often impractical for large-scale mapping. Remote sensing provides a promising alternative, enabling the collection of environmental data in a non-intrusive and cost-effective manner (Fassnacht et al. 2016). Some studies have utilized multispectral imagery, such as Landsat 8 and Sentinel-2, to map tree distributions (Kganyago et al. 2018; Masemola et al. 2020; Muthoka et al. 2021). Considering that hyperspectral images (HSIs) can capture subtle spectral differences between tree species (Sankey et al. 2017) and Light Detection and Ranging (LiDAR) provides detailed structural information (Coops et al. 2007; Shcherbacheva et al. 2024), numerous studies have integrated these complementary data sources for tree species mapping (Chen et al. 2024; Liu et al. 2017; Qin et al. 2022; Shen and Cao 2017).

Numerous nectar-producing tree species worldwide have been investigated through remote sensing. For instance, Atanasov et al. (2024) successfully detected flowering Robinia pseudoacacia in European mixed forests using unmanned aerial vehicle (UAV)-based high-resolution imagery. Similarly, UAV-based seasonal imagery at varying spatial resolutions has been employed to identify *Prosopis glandulosa* shrubs in North America (Jackson et al., 2020). In addition, multispectral satellite imagery has been applied to detect *Melaleuca quinquenervia* in the Florida Everglades (Fuller, 2005). Remote sensing has been also utilized to map *Senegalia mellifera*, *Vachellia tortilis*, and *Commiphora africana* in African savanna regions, although species-level discrimination remains challenging. While recent studies have achieved promising results in mapping *Senegalia mellifera* using hyperspectral and multitemporal imagery (Harkort et al. 2025; Karakizi et al. 2024), these approaches require labels for multiple tree species during model training. Moreover, *Vachellia tortilis* and *Commiphora africana* are seldom distinguished as individual classes. They are frequently grouped with morphologically similar taxa, such as *Vachellia* spp. or *Acacia-Commiphora* woodland types (Coe 1978; Kamau 2017; Karakizi et al. 2024).

Multiclass classification approaches have been extensively applied to tree species mapping tasks (Apostol et al. 2020; Chen et al. 2024; Mäyrä et al. 2021; Qin et al. 2022) and crop mapping (Tang et al. 2024; Wang et al. 2024). Such methods typically require extensive labeled datasets encompassing multiple species for classifier training. However, collecting comprehensive labeled data remains highly impractical in regions with rugged terrain, limited infrastructure, and high species diversity (Schäfer et al. 2016). Furthermore, multiclass classification models may become less reliable when many classes exhibit similar spectral signatures and when class distributions are strongly imbalanced, as separating spectrally close and relatively rare target classes from a heterogeneous background is particularly challenging (Zhao et al. 2023b).

One-class classification (OCC) methods aim to detect a single target class without the need for extensive training data from other classes, offering a practical solution for tree species mapping (Műnoz-Marí et al. 2010; Ruff et al. 2018). OCC methods are generally categorized into positive-only classifiers (P classifiers) and positive-unlabeled classifiers (PU classifiers). P classifiers are trained exclusively on positive samples without relying on negative or unlabeled data, such as one-class support vector machine (Schölkopf et al. 1999), support vector data description (Sanchez-Hernandez et al. 2007), isolation forest (Liu et al. 2008), and maximum entropy (Maxent) (Phillips et al. 2004). Maxent estimates the probability distribution of species occurrence by selecting the most uniform distribution which satisfies a set of environmental constraints.

Considering that incorporating unlabeled data can further improve model performance (Mack and Waske 2017), PU classifiers utilize both positive and unlabeled samples during training. PU-classifiers include a variety of approaches, such as the biased support vector machine (BSVM) (Liu et al. 2003; Piiroinen et al. 2018), the positive and unlabeled learning framework (Elkan and Noto 2008; Li et al. 2010) and the presence and background learning algorithm (Li et al. 2011), hierarchical one-class detection with background (HOCD) (Chang et al. 2024), unbiased risk estimation methods (Du Plessis et al. 2015; Lei et al. 2021; Zhao et al. 2022), one-class risk estimation techniques (Zhao et al. 2023b), and class prior-free methods (Zhao et al. 2023a). The traditional method BSVM (Liu et al. 2003) aims to establish a decision boundary that distinguishes the majority of data points from outliers using positive and unlabeled samples. Based on this, BSVM with minimum distance (hereafter referred to as BSVM) introduces improved strategies for parameter selection and threshold determination (Piiroinen et al. 2018). In addition, HOCD (Chang et al. 2024) derives class signatures from labeled target samples while treating all remaining pixels as unlabeled background, and performs multiclass mapping by detecting one class at a time in a prioritized hierarchical tree.

The deep one-class crop framework (DOCC) (Lei et al. 2021) and the invasive tree species detection framework (ITreeDet) (Zhao et al. 2022) are both built upon unbiased risk estimation (Du Plessis et al. 2015). These methods improve negative class risk estimation in deep feature learning to ensure non-negative loss values. The hyperspectral one-class risk estimation method (HOneCls) (Zhao et al. 2023b) is a patch-free deep learning OCC method that incorporates novel strategies to mitigate overfitting and address distribution imbalance. The Taylor series expansion-based variational framework (T-HOneCls) (Zhao et al. 2023a) has been designed for class prior-free deep PU learning in HSI, introducing a Taylor variational loss to eliminate the need for class prior estimation.

Although existing classification methods can achieve satisfactory accuracy within their trained domains, their performance often degrades when applied to untrained domains due to the discrepancies between source and target data distributions (Tuia et al. 2011), commonly referred to as domain shift. Domain shift typically arises from differences in imaging conditions, atmospheric states, vegetation phenology, or geographic variability in species distribution (Ma et al. 2024). Numerous transfer learning approaches have been developed to address domain shift in remote sensing classification. The methods utilizing labeled target data can be broadly categorized into fine-tuning-based transfer learning (Abdalla et al. 2019), multi-task learning (Sun et al. 2022), and few-shot learning (Cheng et al. 2023). In the absence of labeled target samples, unsupervised domain adaptation (UDA) (Feng et al. 2024; Gogoll et al. 2020; Kwak and Park 2022; Wang et al. 2023) has been proposed to align feature distributions between source and target domains. In addition, self-supervised learning (Sun et al. 2022) introduces supervised information within target domains, thereby eliminating the dependence on labeled samples. For OCC models, one-class support vector machine has been integrated into fine-tuning-based transfer learning frameworks to reduce reliance on negative labeled samples (Chen and Liu 2014; Xue and Beauseroy 2017). Adversarial learning has also been employed to align feature distributions across domains during model training (Chi and Mao 2024; Ding et al. 2023; Mao et al. 2023). However, most existing domain adaptation approaches for hyperspectral classification still assume either access to a small set of labeled samples in the target domain, i.e., OCC models with supervised domain adaptation or comprehensive multiclass annotations in the source domain, i.e., multiclass classifiers with UDA. For one-class classifiers with supervised domain adaptation, the key limitation is that a small number of labeled target positive data is required to guide the adaptation process. However, in practical mapping, collecting reliable labels in new domains is often costly, thereby limiting their applicability. For multiclass classifiers with UDA, adapting them to one-class mapping introduces additional challenges. Besides the need for comprehensive multiclass source labels, the issues of spectral similarity and class imbalance could become more pronounced in cross-domain settings. Domain shift further changes the class-conditional distributions. Limited research has addressed the case where target domains completely lack labeled data, although this is a common and critical challenge in practical large-scale mapping applications. Improving the generalization capacity of OCC models in label-scarce target domains remains to be explored.

This study proposes a hyperspectral unsupervised domain adaptation OCC framework (HyUDA-One) to enhance the generalization capability for tree species mapping. Taita-Taveta County in southeastern Kenya was selected as a representative study region encompassing arid lowland savanna ecosystems. The main contributions of this study are outlined as follows: (1) A positive–unlabeled one-class domain adaptation framework with the spatial–spectral regularized pseudo-positive learning (SSPPL) was introduced to mitigate domain shift and improve generalizability to unlabeled target domains. (2) The three key nectar-producing tree species in savanna landscapes, i.e., *Senegalia mellifera*, *Vachellia tortilis* and *Commiphora africana*, were individually mapped at the species level for the first time using the proposed HyUDA-One framework, trained exclusively with positive and unlabeled samples. (3) Spatial distributions of nectar source availability were derived from the tree species density maps, providing decision support for the economically and environmentally sustainable development of beekeeping in savanna landscapes in Kenya and similar regions worldwide.

## 2. Materials

### 2.1. Study site

This study focuses on Taita-Taveta County, located in southeastern Kenya (Fig. 1a). The study sites include the Choke-Kutima ranch (3° 25′ S, 38° 18′ E) and Lumo Community Wildlife Sanctuary (3° 27′ S, 38° 15′ E). Choke is predominantly covered by bushland (Fig. 1b), with an area of 49 km$^2$ (7 km × 7 km), while Lumo is mainly characterized by grassland (Fig. 1c), spanning 81 km$^2$ (9 km × 9 km). Situated within a lowland savanna ecosystem (Figs. 2a and 2b), both sites hold ecological significance for conservation and potential for apiculture. The landscapes comprise a variety of vegetation types, with key tree species for beekeeping including *Senegalia mellifera*, *Vachellia tortilis*, and

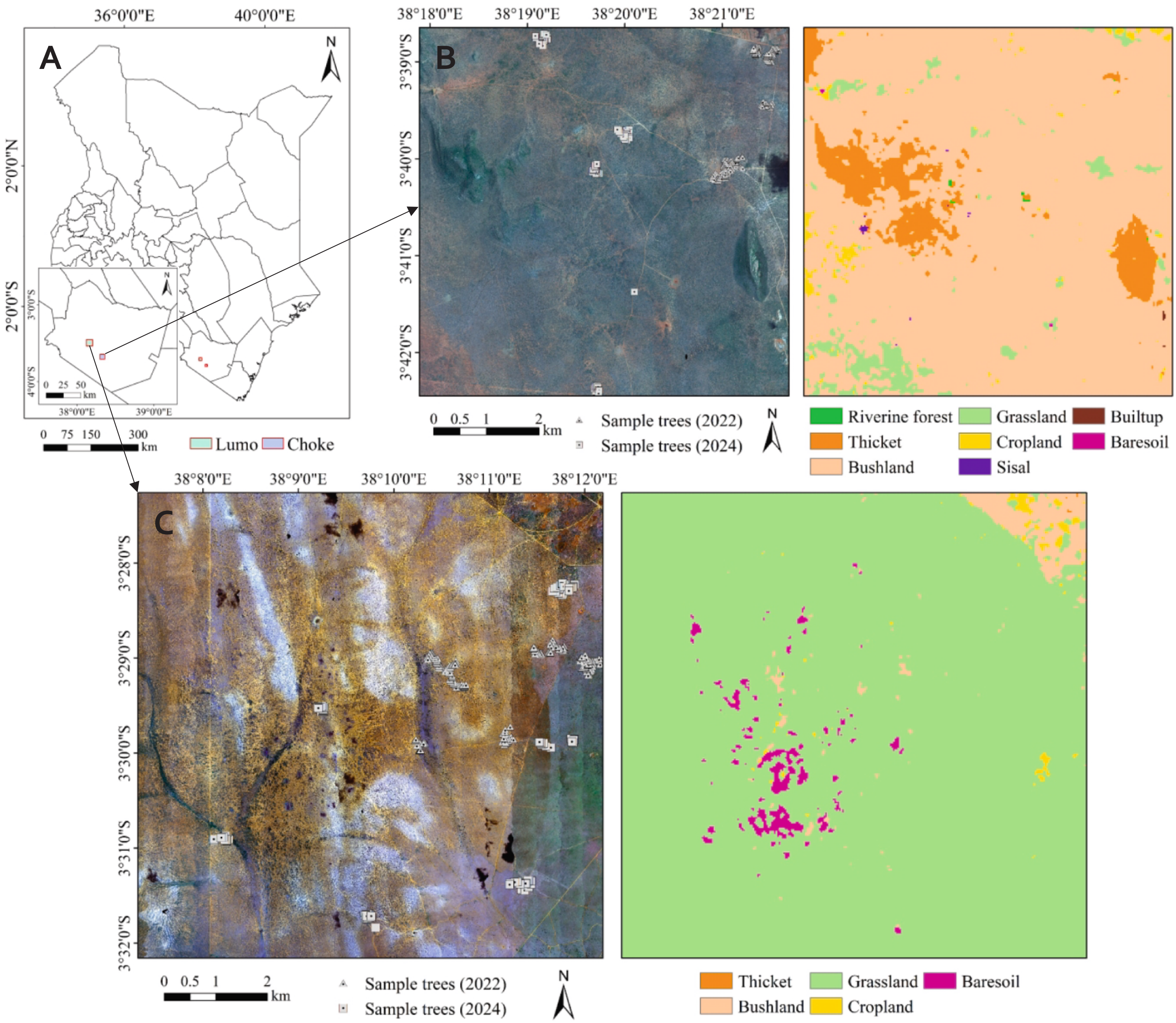

**Fig. 1.** (a) The study sites in the southeast of Kenya. (b) Hyperspectral imagery mosaic of the Choke-Kutima ranchand and the land cover map of Choke in 2020. (c) Hyperspectral imagery mosaic of Lumo Community Wildlife Sanctuaryand and the corresponding land cover map in 2020. Field measurement locations are overlaid on hyperspectral imagery.

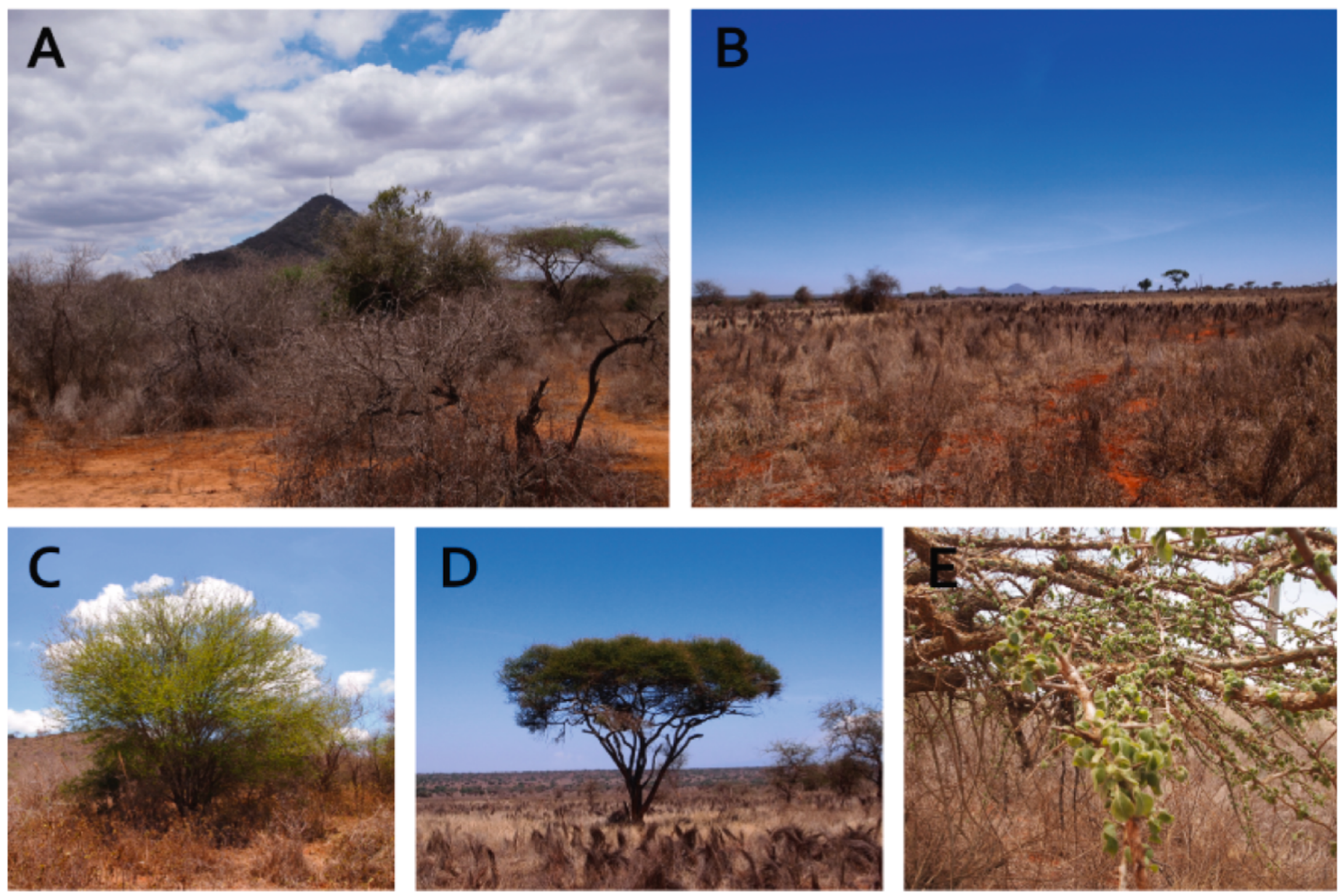

**Fig. 2.** (a) *Acacia-Commiphora* bushland in Choke during the dry season. (b) Open savanna landscapes with scattered trees in Lumo. (c) *Senegalia mellifera* growing along a roadside. (d) *Vachellia tortilis* with distinctive umbrella-shaped canopy. (e) Close-up of the small, rounded leaves of *Commiphora africana*. (Photos: P. Mwasi, 2024).

*Commiphora africana*. Other commonly occurring species include *Albizia amara* and *Balanites maughamii*, while additional species such as *Balanites aegyptiaca*, *Boscia angustifolia*, and *Cordia monoica* further contribute to the local floristic diversity.

*Senegalia mellifera*, *Vachellia tortilis*, and *Commiphora africana* were selected as target tree species in this study due to their significant value for apiculture. *Senegalia mellifera* and *Vachellia tortilis* are two key nectar sources for honeybees (Figs. 2c and 2d). Both species are well adapted to the climatic conditions of savanna ecosystems. *Senegalia mellifera* is not palatable to most herbivores due to its multi-stemmed and thorny thicket structure (Karakizi et al. 2024). *Vachellia tortilis*, characterized by its distinctive umbrella-shaped canopy, also serves as an important fodder source for livestock (Yadeta et al. 2018). *Commiphora africana* is a deciduous tree species native to dry tropical regions (Fig. 2e). Notable for its aromatic resin, it functions as a valuable nectar source for stingless bees (Mduda et al. 2023).

### 2.2. Remote sensing data

Hyperspectral data was collected using an AisaKestrel 10 sensor with a 39° field of view. The flight heights were approximately 1400 m for Choke and 890 m for Lumo. Data acquisition took place in March 2022 for Choke and in February 2022 for Lumo. The resulting imagery had a spatial resolution of 1 m and provided detailed spectral information across 92 bands, covering the wavelength range of 381–1003 nm, with a full width at half maximum (FWHM) of 6.8 nm. A series of preprocessing steps were applied to the hyperspectral imagery. Radiometric and orthorectification corrections were performed using CaliGeoPro 2.2 software. Subsequently, atmospheric and bidirectional reflectance distribution function (BRDF) corrections were performed using DROACOR 2.0 software. The hyperspectral image mosaics of Choke and Lumo are presented in Figs. 1b and 1c. Fig. 3 illustrates noticeable differences in the average spectral signatures of the same tree species, e.g., *Vachellia tortilis*, between Choke and Lumo.

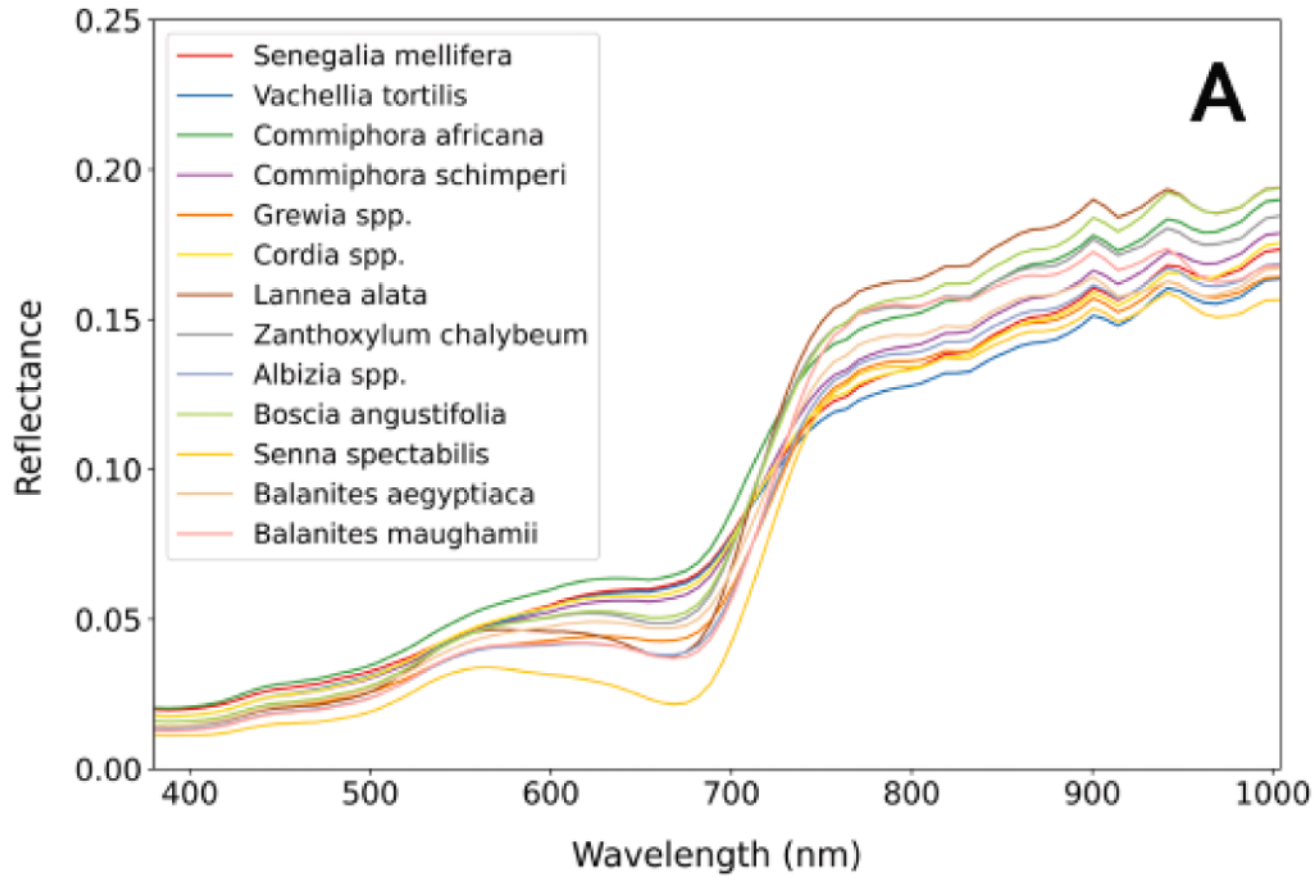

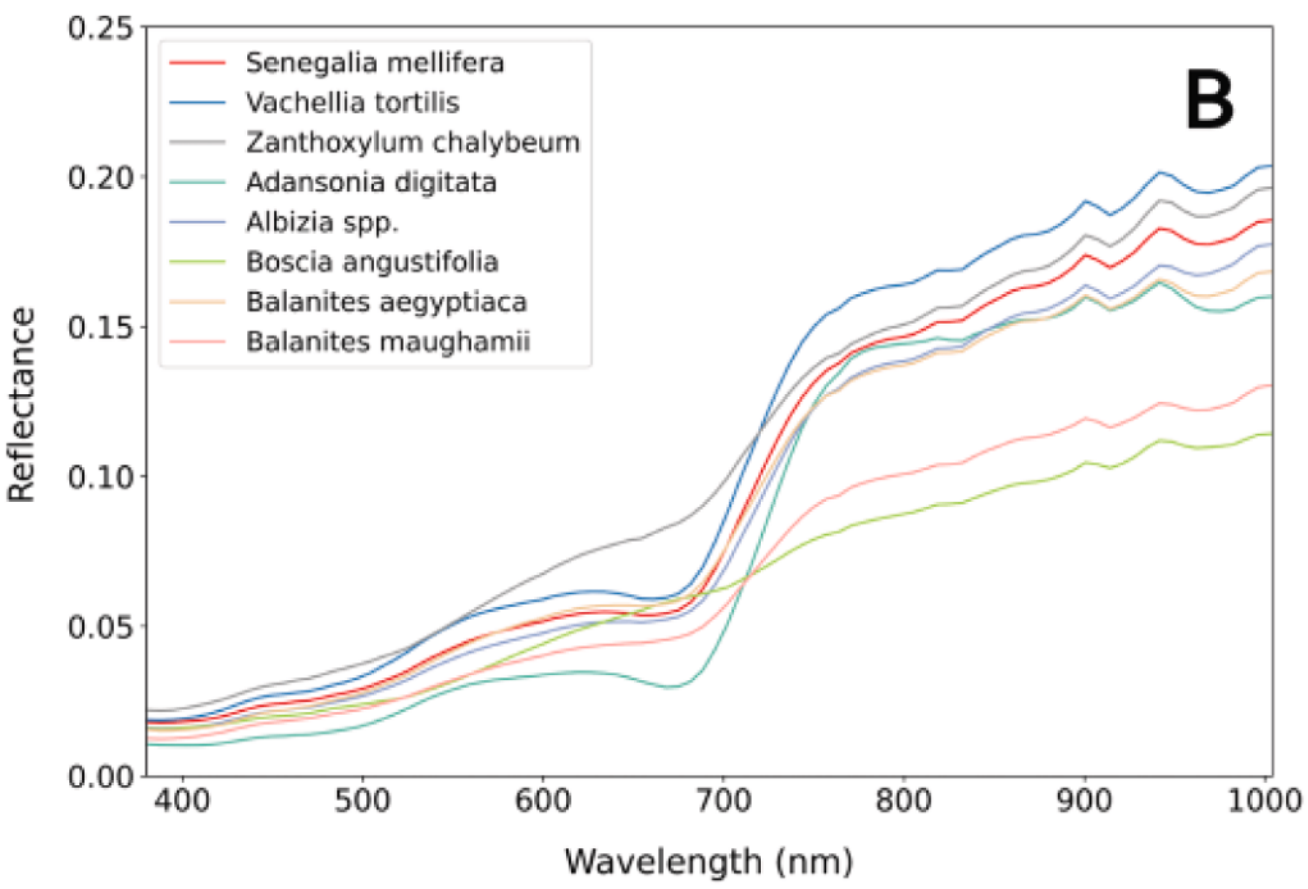

**Fig. 3.** (a) Visualization of the average spectra derived from hyperspectral imagery for tree species in Choke. (b) Average spectral profiles of tree species in Lumo from hyperspectral imagery.

LiDAR data was acquired in parallel with the hyperspectral data using a Leica ALS60 sensor. The data was essential for capturing the three-dimensional structure of vegetation. The average LiDAR point density was 2.41 points/m$^2$ for Choke and 4.17 points/m$^2$ for Lumo. The LiDAR data was processed using LAStools software to generate a canopy height model (CHM), which was subsequently utilized to generate a tree mask. LiDAR-derived features were not directly used as inputs to the HyUDA-One model.

### 2.3. Reference data

Field data was collected to provide ground reference information for model training and validation. Tree-covered areas with heights exceeding 1.5 m were identified using the CHM maps. In combination with accessible roads in rugged terrain, sampling transects were established within the identified areas. Sampling was conducted within 500-m buffers along the transects to record trees with various species, crown sizes and growth conditions. The positions of the trees were measured using a GNSS receiver (Trimble GeoXH 6000, Trimble Inc., Sunnyvale, CA, USA), and differential correction was applied using a base station (Trimble NetR9). A total of 830 trees were recorded in December 2022, followed by 274 in September 2024. High-resolution airborne (RGB-NIR) imagery, acquired simultaneously with the hyperspectral data, was utilized to assist in the labeling and identification of tree species. After excluding the ground measurements that did not match the hyperspectral data, the final dataset comprised 670 tree crowns, corresponding to 18,586 pixels. In addition to the three melliferous tree species, the dataset also included other encountered tree species. These comprised *Albizia amara*, *Balanites maughamii*, *Balanites aegyptiaca*, *Boscia angustifolia*, *Cordia monoica*, *Grewia bicolor* and *Albizia anthelmintica*.

## 3. Methods

A novel HyUDA-One framework is proposed for tree species mapping by introducing an unsupervised iterative adaptation strategy for unlabeled target domains. As illustrated in Fig. 4, the proposed framework comprises two stages: 1) training on source domains using a PU risk estimator, and 2) unsupervised iterative adaptation on target domains. The details are described in the following sections.

### 3.1. Dimensionality reduction

Considering the high dimensionality and spectral redundancy of hyperspectral data, the minimum noise fraction (MNF) transformation (Green et al. 1988) was employed to reduce redundant information and improve signal quality (Shi and Pun 2019). MNF can be viewed as a two-step process that integrates noise whitening and principal component analysis (PCA), where PCA is performed on the noise-whitened data to rank components based on their signal-to-noise ratio. The first 15 MNF-transformed bands were selected for subsequent processing based on their eigenvalues and visual assessment of image quality.

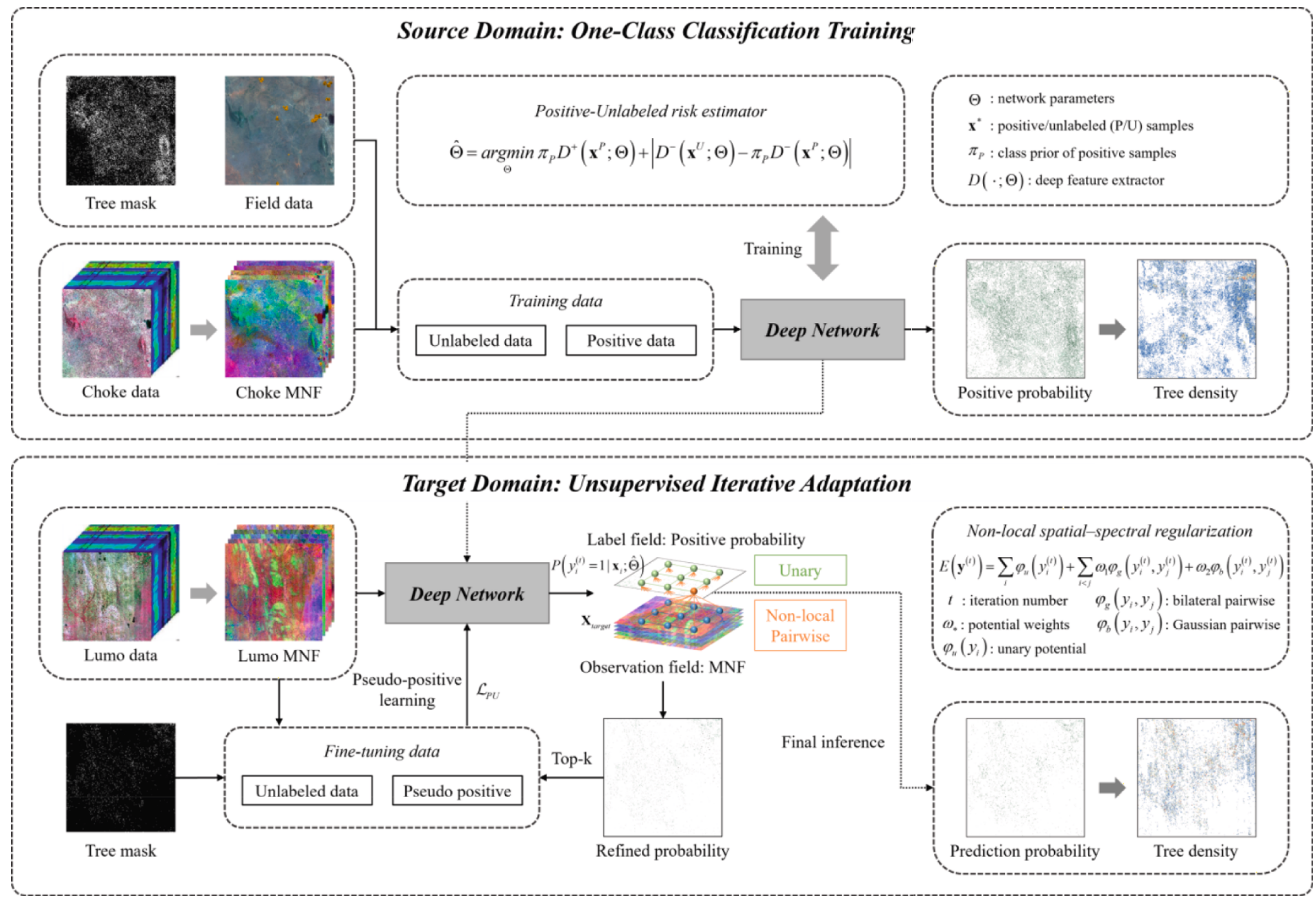

**Fig. 4.** Overview of the proposed hyperspectral unsupervised domain adaptation one-class classification framework (HyUDA-One) for tree species mapping. In the source domain (top), a deep one-class classifier is trained using positive (P) and unlabeled (U) samples with a PU risk estimator to produce species probability maps. In the target domain (bottom), the pretrained network is iteratively fine-tuned on unlabeled target data through the spatial–spectral regularized pseudo-positive learning (SSPPL), leading to refined prediction probability maps.

### 3.2. Tree mask generation

To mitigate the influence of non-tree pixels on classification accuracy (Piiroinen et al. 2018), a tree mask was generated using the hyperspectral data and the LiDAR CHM. Specifically, the near-infrared (NIR) reflectance at 840 nm and the normalized difference vegetation index (NDVI, computed using NIR at 840 nm and the red band at 661 nm) were extracted from the hyperspectral imagery. The pixels with NIR reflectance below 0.2, NDVI below 0.3, or CHM below 1.5 m were excluded to eliminate non-vegetative areas and low-lying vegetation from subsequent classification.

### 3.3. Unsupervised domain adaptation for one-class classification

#### 3.3.1. Source domain training with PU loss

Based on the MNF components derived from the hyperspectral data and the tree mask, a subset of positive samples was extracted from the Choke field dataset for model training. In addition, a large number of randomly selected unlabeled samples were incorporated into the training process. A deep neural network was trained using positive and unlabeled samples, guided by the PU risk estimator illustrated in Fig. 4, to distinguish the target tree species. The trained network was then utilized to predict the presence of positive tree species in Choke and to generate species-specific density maps. Details of the PU risk estimator are provided below.

Traditional supervised classification methods typically rely on binary loss functions, which require labeled datasets containing both positive and negative samples. The objective of such binary classification can be formulated as follows:

$$\widehat{\Theta} = \underset{\Theta}{argmin}\,\pi_P D^+\left(\boldsymbol{x}^P;\Theta\right) + (1-\pi_P)D^-\left(\boldsymbol{x}^N;\Theta\right) \tag{1}$$

where $\Theta$ denotes the parameters of the deep network, $\pi_P$ is the class prior of positive samples, and $\boldsymbol{x}^P$ and $\boldsymbol{x}^N$ represent positive and negative samples from the source MNF components $\boldsymbol{X}_{source}$, respectively. $D^+\left(\boldsymbol{x}^P;\Theta\right) = \frac{1}{n_P}\sum_{i=1}^{n_P}\mathscr{L}\left(D\left(\boldsymbol{x}_i^P;\Theta\right),+1\right)$ and $D^-\left(\boldsymbol{x}^P;\Theta\right) = \frac{1}{n_N}\sum_{j=1}^{n_N}\mathscr{L}\left(D\left(\boldsymbol{x}^N;\Theta\right),-1\right)$ refer to the average loss of the positive and negative classes, where $\mathscr{L}[\cdot]$ denotes to the selected loss function.

Recording comprehensive information on all non-target species as negative samples is particularly challenging in species-rich environments, especially when only a single or a few tree species are of interest. In such cases, OCC methods offer a practical alternative. Among these, PU learning strategies utilize abundant unlabeled samples to implicitly infer negative information. Considering the risk of the negative samples can be estimated by $D^-\left(\boldsymbol{x}^N;\Theta\right) = \left(D^-\left(\boldsymbol{x}^U;\Theta\right) - \pi_P D^-\left(\boldsymbol{x}^P;\Theta\right)\right)/(1-\pi_P)$ (Du Plessis et al. 2015), the unbiased PU risk estimator can be calculated as:

$$\widehat{\Theta} = \underset{\Theta}{argmin}\,\pi_P D^+\left(\boldsymbol{x}^P;\Theta\right) + D^-\left(\boldsymbol{x}^U;\Theta\right) - \pi_P D^-\left(\boldsymbol{x}^P;\Theta\right) \tag{2}$$

where $\boldsymbol{x}^U$ denotes unlabeled samples, which are randomly selected from all the tree pixels. Therein, $D^-\left(\boldsymbol{x}^U;\Theta\right) = \frac{1}{n_U}\sum_{j=1}^{n_U}\mathscr{L}\left(D\left(\boldsymbol{x}_j^U;\Theta\right),-1\right)$ and $D^-\left(\boldsymbol{x}^P;\Theta\right) = \frac{1}{n_P}\sum_{i=1}^{n_P}\mathscr{L}\left(D\left(\boldsymbol{x}_i^P;\Theta\right),-1\right)$. However, this formulation may suffer from instability and overfitting issues, since the negative loss $D^-\left(\boldsymbol{x}^N;\Theta\right)$ should not be less than zero (Lei et al. 2021). To improve model robustness and stable convergence during the initial supervised training stage, a modified risk estimator (Zhao et al. 2022) is adopted. An absolute value operation is incorporated into the unbiased PU risk estimator as follows:

$$\widehat{\Theta} = \underset{\Theta}{argmin}\,\pi_P D^+\left(\boldsymbol{x}^P;\Theta\right) + \left|D^-\left(\boldsymbol{x}^U;\Theta\right) - \pi_P D^-\left(\boldsymbol{x}^P;\Theta\right)\right| \tag{3}$$

#### 3.3.2. Unsupervised iterative adaptation

To improve the generalization capability of OCC models, this study proposes an unsupervised iterative adaptation strategy designed for the positive–unlabeled one-class setting in unlabeled target domains. First, a PU-based one-class classifier was trained on the source domain. The resulting positive probability maps in the target domain were then refined through a non-local spatial–spectral regularization with dense conditional random field (CRF) (Krähenbühl and Koltun, 2011). High-confidence pixels selected from the refined probabilities were treated as pseudo-positive samples in the target domain. These samples were then used to update the PU risk estimator and fine-tune the network parameters. The spatial–spectral regularized pseudo-positive learning was repeated iteratively, gradually adapting the one-class classifier to the target domain without requiring any labeled target samples.

Specifically, the model trained in the first stage was directly applied to the MNF data $\boldsymbol{X}_{target}$ from the target domain to generate the positive predictions $P\left(y_i = 1|\boldsymbol{x}_i;\widehat{\Theta}\right)$. For each pixel $i$, the predicted probabilities at iteration $t$ are expressed as:

$$P\left(y_i^{(t)} = 1|\boldsymbol{x}_i;\widehat{\Theta}^{(t)}\right) = D\left(\boldsymbol{x}_i;\widehat{\Theta}^{(t)}\right) \tag{4}$$

where $\boldsymbol{x}_i$ denotes the feature patch of the MNF components at pixel $i$, and $\widehat{\Theta}^{(t)}$ represents the network parameters at iteration $t$. The corresponding negative probability is computed as $P\left(y_i^{(t)} = -1|\boldsymbol{x}_i;\widehat{\Theta}^{(t)}\right) = 1 - P\left(y_i^{(t)} = 1|\boldsymbol{x}_i;\widehat{\Theta}^{(t)}\right)$. Afterwards, the non-local spatial–spectral regularization was performed on the probabilities, integrating the unary potentials derived from the model outputs with pairwise potentials that encode both spatial and spectral context. The non-local spatial–spectral regularization with dense CRF is formulated as:

$$E\left(\boldsymbol{y}^{(t)}\right) = \sum_i \varphi_u\left(y_i^{(t)}\right) + \sum_{i<j}\omega_1\varphi_g\left(y_i^{(t)},y_j^{(t)}\right) + \omega_2\varphi_b\left(y_i^{(t)},y_j^{(t)}\right) \tag{5}$$

where $\varphi_u\left(y_i^{(t)}\right)$ represents the unary potential, $\varphi_g\left(y_i^{(t)},y_j^{(t)}\right)$ denotes the Gaussian pairwise potential, and $\varphi_b\left(y_i^{(t)},y_j^{(t)}\right)$ refers to the bilateral pairwise potential. $\omega_1$ and $\omega_2$ denote the weights for the Gaussian and bilateral potentials. The unary potential $\varphi_u\left(y_i^{(t)}\right)$ is calculated based on the negative log-likelihood of the probabilities as follows:

$$\varphi_u\left(y_i^{(t)}\right) = -\log\left(P\left(y_i^{(t)}|\boldsymbol{x}_i;\widehat{\Theta}^{(t)}\right)\right) \tag{6}$$

The Gaussian pairwise potential $\varphi_g\left(y_i^{(t)},y_j^{(t)}\right)$ is aimed at considering spatial similarity, calculated as follows:

$$\varphi_g\left(y_i^{(t)},y_j^{(t)}\right) = \mu\left(y_i^{(t)},y_j^{(t)}\right)\exp\left(-\frac{\left\|\boldsymbol{p}_i - \boldsymbol{p}_j\right\|^2}{2\sigma_\alpha}\right) \tag{7}$$

where $\boldsymbol{p}_i$ and $\boldsymbol{p}_j$ denote spatial coordinates of pixels $i$ and $j$. The compatibility function is defined using the Potts model $\mu\left(y_i^{(t)},y_j^{(t)}\right) = \mathbb{I}\left(y_i^{(t)} \neq y_j^{(t)}\right)$. $\sigma_\alpha$ controls the distance smoothness. The bilateral pairwise potential $\varphi_b\left(y_i^{(t)},y_j^{(t)}\right)$ aims to capture spatial-spectral correlation as follows:

$$\varphi_b\left(y_i^{(t)}, y_j^{(t)}\right) = \mu\left(y_i^{(t)}, y_j^{(t)}\right)\exp\left(-\frac{\left\|\boldsymbol{p}_i - \boldsymbol{p}_j\right\|^2}{2\sigma_\beta} - \frac{\left\|\boldsymbol{x}_i - \boldsymbol{x}_j\right\|^2}{2\sigma_\gamma}\right) \tag{8}$$

where $\sigma_\beta$ and $\sigma_\gamma$ control the spatial and spectral smoothness, respectively. Subsequently, the refined probabilities $P^{CRF}\left(y_i^{(t)} = 1|\boldsymbol{x}_i; \widehat{\Theta}^{(t)}\right)$ were generated by minimizing the energy function. The top-$k$ pixels with the highest refined probabilities were selected as pseudo-positive samples to iteratively fine-tune the model. The selection is described by:

$$\widehat{y}_i^{(t)} = \begin{cases} 1, & \text{if } P^{CRF}\left(y_i^{(t)} = 1|\boldsymbol{x}_i; \widehat{\Theta}^{(t)}\right) \geqslant \tau^{(t)} \\ 0, & \text{otherwise} \end{cases} \tag{9}$$

where the adaptive threshold $\tau^{(t)}$ was determined based on the top-$k$ strategy during iteration. With these pseudo-positive samples, the network parameters $\widehat{\Theta}^{(t)}$ were iteratively updated using the PU risk estimator in equation (3) as follows:

$$\widehat{\Theta}^{(t+1)} = \underset{\widehat{\Theta}^{(t)}}{argmin}\,\pi_P^{(t)} D^+\left(\boldsymbol{x}_{pseudo}^{(t)}; \widehat{\Theta}^{(t)}\right) + \left| D^-\left(\boldsymbol{x}_{target}^{U}; \widehat{\Theta}^{(t)}\right) - \pi_P^{(t)} D^-\left(\boldsymbol{x}_{pseudo}^{(t)}; \widehat{\Theta}^{(t)}\right) \right| \tag{10}$$

**Algorithm 1** Unsupervised iterative adaptation via the spatial–spectral regularized pseudo-positive learning

**Input:**
The CNN parameters $\widehat{\Theta}^{(0)}$ trained from the source domain
The MNF data in the target regions $\boldsymbol{X}_{target}$
Non-local spatial–spectral regularization parameters: $\omega_1$, $\omega_2$, $\sigma_\alpha$, $\sigma_\beta$, $\sigma_\gamma$
Hyperparameters: Number of iterations $T$, top-$k$ value

**Output:**
Adapted CNN parameters $\widehat{\Theta}^{(T)}$

1: Load CNN parameters $\widehat{\Theta}^{(0)}$ to the deep network
2: **for** $t = 0, 1, 2, \ldots, T-1$ **do**
3: Generate predictions $P\left(y_i^{(t)} = 1|\boldsymbol{x}_i; \widehat{\Theta}^{(t)}\right)$ by Eq (4)
4: Non-local spatial–spectral regularization with dense CRF (mean-field): compute unary potentials $\varphi_u\left(y_i^{(t)}\right)$, Gaussian pairwise potentials $\varphi_g\left(y_i^{(t)}, y_j^{(t)}\right)$ and bilateral pairwise potentials $\varphi_b\left(y_i^{(t)}, y_j^{(t)}\right)$ for each pixel $i$ by Eq (6), 7, 8
5: Initialize $Q_i\left(y_i^{(t)}\right) \propto \exp\left(-\varphi_u\left(y_i^{(t)}\right)\right)$
6: **for** $l = 1, 2, \ldots, n_{iter}$ **do**
7: $Q_i^{(l)}\left(y_i^{(t)}\right) \propto \exp\left(-\varphi_u\left(y_i^{(t)}\right) - \sum_j \omega_1 \varphi_g\left(y_i^{(t)}, y_j^{(t)}\right) Q_j^{(l-1)}\left(y_j^{(t)}\right) - \omega_2 \varphi_b\left(y_i^{(t)}, y_j^{(t)}\right) Q_j^{(l-1)}\left(y_j^{(t)}\right)\right)$
8: **end for**
9: Update probabilities $P^{CRF}\left(y_i^{(t)} = 1|\boldsymbol{x}_i; \widehat{\Theta}^{(t)}\right) = Q_i^{(n_{iter})}\left(y_i^{(t)}\right)$
10: Calculate adaptive threshold $\tau^{(t)}$ based on $k$
11: Generate pseudo labels by Eq (9)
12: Estimate class prior $\pi_P^{(t)}$
13: Fine-tuning network parameters $\widehat{\Theta}^{(t)}$ by Eq (10)
14: Set $\widehat{\Theta}^{(t+1)} \leftarrow \widehat{\Theta}^{(t)}$
15: **end for**
16: Return final adapted CNN parameters $\widehat{\Theta}^{(T)}$ and the final probability map $P\left(y_i^{(T)} = 1|\boldsymbol{x}_i; \widehat{\Theta}^{(T)}\right)$

where $\boldsymbol{x}_{pseudo}^{(t)}$ denotes the pseudo-positive samples generated at iteration $t$ and $\boldsymbol{x}_{target}^{U}$ represents unlabeled samples in the target domain. The class prior $\pi_P^{(t)}$ was estimated based on pseudo labels. The entire process is outlined in Algorithm 1. A mean-field approximation was employed to minimize the energy function $E\left(\boldsymbol{y}^{(t)}\right)$ and compute refined probabilities, as described in the steps 5 to 9.

### 3.4. Comparison to existing one-class and multiclass classifiers

Seven OCC classifiers were compared to map melliferous tree species in Choke. Maxent (Phillips et al. 2004) was selected as a representative P classifier. The PU classifiers included several state-of-the-art approaches: BSVM (Piiroinen et al. 2018), DOCC (Lei et al. 2021), ITreeDet (Zhao et al. 2022), HOneCls (Zhao et al. 2023b), T-HOneCls (Zhao et al. 2023a), and HOCD (Chang et al. 2024). In the Choke experiment, HyUDA-One adopts the same risk estimator and backbone architecture as ITreeDet. When the UDA components are inactive, HyUDA-One reduces to the same one-class classifier as ITreeDet, leading to equivalent performance in Choke. To assess the generalization capability of the above models and validate the effectiveness of the proposed HyUDA-One framework, a generalizability analysis was conducted using the field data from Lumo. All the models were trained on the data from Choke without any pretraining on external data and evaluated on the Lumo dataset. The comparison between multiclass and one-class classifiers was conducted by additionally including a deep multiclass baseline, i.e., fast patch-free global learning (FPGA) (Zheng et al. 2020), and the multilevel unsupervised domain adaptation (MLUDA) framework (Cai et al. 2024). The experiments were conducted in both the source domain Choke and the target domain Lumo. To ensure a fair comparison with the one-class setting, unlabeled samples were treated as negative in the multiclass models.

### 3.5. Validation

In this study, 70 % of the positive samples from Choke were used for model training, while the remaining 30 %, together with all negative samples, were used for validation (Table 1). Individual one-class models were trained and validated for each of the three melliferous tree species. In Lumo, 70 % of the positive samples were used as the training set in the Train: Lumo scenario of the ablation analysis. The remaining 30 % of the positive samples, together with all negative samples from Lumo, were used to assess model performance in both the generalizability analysis and the ablation analysis. To ensure independence between training and testing, data was divided at the crown level to avoid including samples from the same tree. During training, 20,000 unlabeled samples were randomly selected from the masked tree pixels. It is worth noting that only a few *Commiphora africana* were found in Lumo. Therefore, it was excluded in the generalizability analysis.

The quantitative evaluation included precision, recall, the F1-score, and the metrics computed from the receiver operating characteristic (ROC) curve and the precision-recall (PR) curve. Precision measures the proportion of correctly identified positive samples among all samples predicted as positive by the model. Recall reflects the model ability to correctly identify all actual positive samples. The F1-score combines precision and recall into their harmonic mean, calculated as $F1 = (2 \times Presicion \times Recall)/(Presicion + Recall)$. The ROC curve illustrates the trade-off between true positive rate and false positive rate across various thresholds. The area under the ROC curve (AUC) quantifies overall model performance, with higher values indicating better separability between positive and negative classes. The PR curve depicts the trade-off between precision and recall across various thresholds. The average precision (AP), calculated as the area under the PR curve, provides a single metric summarizing the model ability to balance precision and recall. A higher AP value indicates better overall performance in accurately identifying the target class.

An ablation analysis was conducted to assess the contributions of each module in the proposed HyUDA-One framework. The evaluation incorporated the quantitative metrics described above. To further evaluate feature discriminability, t-distributed stochastic neighbor embedding (t-SNE) was utilized to visualize deep features extracted from the final hidden layer of the networks with various modules. Five experimental scenarios were considered: 1) Training directly in the target domain (Train: Lumo), which represents the upper-bound performance with access to labeled target samples; 2) Training solely in the source domain (Train: Choke) without adaptation to target domain; 3) Reflectance alignment (RA): RA performs a band-wise linear transformation of the reflectance of the target domain to match the first- and second-order statistics of the source domain. For each spectral band $b$, the mean and standard deviation of reflectance over tree-covered pixels $\left(\mu^{(b)}_{source}, \sigma^{(b)}_{source}\right)$ in the source domain and $\left(\mu^{(b)}_{target}, \sigma^{(b)}_{target}\right)$ in the target domain are computed, where tree-covered pixels are identified using the CHM–NDVI–NIR-based mask without requiring any labeled samples. Afterwards, each target reflectance value $x^{(b)}_{target}$ is transformed as $\widetilde{x}^{(b)}_{target} = \left(x^{(b)}_{target} - \mu^{(b)}_{target}\right) * \sigma^{(b)}_{source}/\sigma^{(b)}_{target} + \mu^{(b)}_{source}$; 4) Using pseudo-positive learning (PPL) alone for iterative adaptation; and (5) Combining RA and PPL (RA + PPL). The full HyUDA-One framework (RA + SSPPL) integrates reflectance alignment, pseudo label learning, and the non-local spatial–spectral regularization.

**Table 1**
Training and validation samples in Choke and Lumo.

| Study site | Tree species | Training | | | Validation | | | |
|---|---|---|---|---|---|---|---|---|
| | | Positive pixels | Positive crowns | Unlabeled pixels | Positive pixels | Positive crowns | Negative pixels | Negative crowns |
| Choke | *Senegalia mellifera* | 860 | 38 | 20,000 | 361 | 17 | 1,727 | 102 |
| | *Vachellia tortilis* | 1,387 | 28 | 20,000 | 588 | 12 | 8,145 | 251 |
| | *Commiphora africana* | 3,361 | 71 | 20,000 | 997 | 31 | 5,557 | 222 |
| Lumo | *Senegalia mellifera* | 588 | 54 | 20,000 | 190 | 24 | 1,635 | 131 |
| | *Vachellia tortilis* | 4,786 | 113 | 20,000 | 2,106 | 49 | 1,635 | 131 |

### 3.6. Analysis of nectar source availability

Nectar source availability was evaluated based on pixel-wise probability maps generated for each target tree species. Separate analyses were conducted for honeybees and stingless bees. *Senegalia mellifera* and *Vachellia tortilis* serve as primary nectar sources for honeybees, while stingless bees make extensive use of *Commiphora africana*. For each bee group, the availability score at a location $(x, y)$ was calculated using a Gaussian–weighted voting filter:

$$Score(x,y) = \sum_{m=-M}^{M} \sum_{n=-N}^{N} f(x-m, y-n) g(m,n) \tag{11}$$

where $f(x,y)$ denotes the density of tree species and $g(m,n) = e^{-\left(m^2+n^2\right)/2\sigma_s^2}/2\pi\sigma^2$ is the isotropic Gaussian kernel with standard deviation $\sigma_s$. $M$ and $N$ define the spatial extent of the voting window. The kernel assigns higher weights to nearby pixels and reduces the influence of more distant ones, producing a smooth and noise–resistant estimate of nectar source availability. Higher scores indicate locations surrounded by richer nectar sources.

### 3.7. Experimental settings

All deep learning models were implemented using Python 3.7 with the PyTorch framework. In the first stage, the model settings and network architecture followed the original setup (Zhao et al. 2022). In the second stage, the number of iterations $T$ was set to 10, and the top-$k$ value was set to 5000 per iteration. The settings of hyperparameters followed $\omega_1 = 4, \omega_2 = 5, \sigma_\alpha = 5, \sigma_\beta = 30, \sigma_\gamma = 8$ empirically. The gradient thresholding algorithm (Ramaswamy et al., 2016) was applied for estimating class priors. All computation was performed on a system equipped with Intel(R) Xeon(R) E5-2690 CPUs (2.60 GHz) and a single NVIDIA Tesla P100 GPU.

## 4. Results

### 4.1. The classification results for the source domain (Choke)

DOCC, ITreeDet and HOCD achieved the best overall performance in the source domain. According to Table 2, ITreeDet achieved the highest F1-score of 0.788 for *Senegalia mellifera*. DOCC obtained slightly higher AUC and AP values of 0.943 and 0.861, respectively, indicating superior discrimination capability across varying thresholds. ITreeDet also yielded the highest precision for mapping *Senegalia mellifera*, while DOCC achieved the highest recall (Fig. 5). For *Vachellia tortilis*, HOCD achieved the highest precision, while ITreeDet outperformed all the other models across all the other metrics. In contrast, Maxent, BSVM, HOneCls and T-HOneCls exhibited unsatisfactory performances in F1-score and AP. For *Commiphora africana*, HOCD achieved the highest F1-score of 0.814 and the best AP value of 0.831. DOCC exhibited the highest AUC of 0.953. Both HOneCls and T-HOneCls also demonstrated robust performance. The remaining OCC classifiers exhibited comparable but slightly lower performance.

### 4.2. Generalization capability to the target domain (Lumo)

The proposed HyUDA-One approach substantially outperformed all the comparative methods across both *Senegalia mellifera* and *Vachellia tortilis* (Tabel 3 and Fig. 6), demonstrating its strong generalization capability. Specifically, for *Senegalia mellifera*, HyUDA-One achieved superior precision and recall values, resulting in an F1-score of 0.756, AUC of 0.890, and AP of 0.796. In contrast, the other methods exhibited relatively poor performance, indicating their limited ability to generalize across landscapes for these two melliferous tree species. For *Vachellia tortilis*, BSVM, DOCC, ITreeDet, and HOCD failed to maintain their original accuracy, with notable declines in both AUC and AP. Maxent, HOneCls and T-HOneCls displayed satisfactory performance, suggesting their potential generalization capability for *Vachellia tortilis*. In comparison, HyUDA-One showed better stability in both precision and recall, leading to superior overall performance with an F1-score of 0.884, AUC of 0.943, and AP of 0.958.

Additionally, the local prediction maps in Figs. 7 and 8 reveal subtle differences in the spatial performance of the methods. For *Senegalia mellifera* (Fig. 7), Maxent, BSVM, ITreeDet, HOneCls, T-HOneCls, and HOCD were able to detect the target species but also misclassified non-target species as positive. DOCC failed to correctly identify the target species. In the case of *Vachellia tortilis* (Fig. 8), all the comparative methods tended to misclassify other tree species as positive. In contrast, HyUDA-One accurately identified the target tree species while effectively minimizing false positives among non-target tree species.

### 4.3. The comparison of multiclass and one-class classifiers

The comparison indicated that the one-class classifiers provide more robust species-specific mapping performance than the multiclass classifiers. In the source domain Choke, the multiclass approaches exhibited weaker and less stable performance across species (Table 4). Although good results were achieved for *Vachellia tortilis*, the performance droped substantially for *Senegalia mellifera* and *Commiphora africana*. For

**Table 2**
F1-score, AUC,and AP of all OCC classifiers for mapping melliferous tree species in Choke. The best performance is marked with bold.

| Method | *Senegalia mellifera* | | | *Vachellia tortilis* | | | *Commiphora africana* | | |
|---|---|---|---|---|---|---|---|---|---|
| | F1-score | AUC | AP | F1-score | AUC | AP | F1-score | AUC | AP |
| Maxent | 0.713 | 0.925 | 0.795 | 0.657 | 0.917 | 0.682 | 0.737 | 0.927 | 0.78 |
| BSVM | 0.743 | 0.916 | 0.795 | 0.587 | 0.945 | 0.626 | 0.729 | 0.932 | 0.801 |
| DOCC | 0.78 | **0.943** | **0.861** | 0.794 | 0.987 | 0.89 | 0.768 | **0.953** | 0.777 |
| ITreeDet (base model) | **0.788** | 0.942 | 0.848 | **0.845** | **0.993** | **0.931** | 0.759 | 0.949 | 0.712 |
| HOneCls | 0.761 | 0.900 | 0.752 | 0.667 | 0.961 | 0.699 | 0.766 | 0.950 | 0.826 |
| T-HOneCls | 0.744 | 0.892 | 0.769 | 0.511 | 0.940 | 0.566 | 0.753 | 0.941 | 0.819 |
| HOCD | 0.785 | 0.940 | 0.816 | 0.772 | 0.919 | 0.840 | **0.814** | 0.874 | **0.831** |

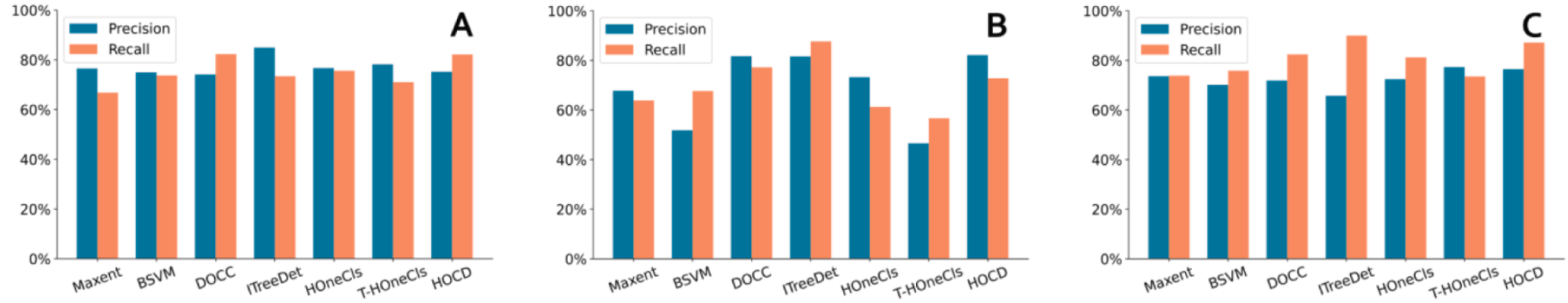

**Fig. 5.** The precision and recall (percentages) of the OCC classifiers for mapping (a) *Senegalia mellifera*, (b) *Vachelia tortillis*, and (c) *Commiphora africana* in Choke.

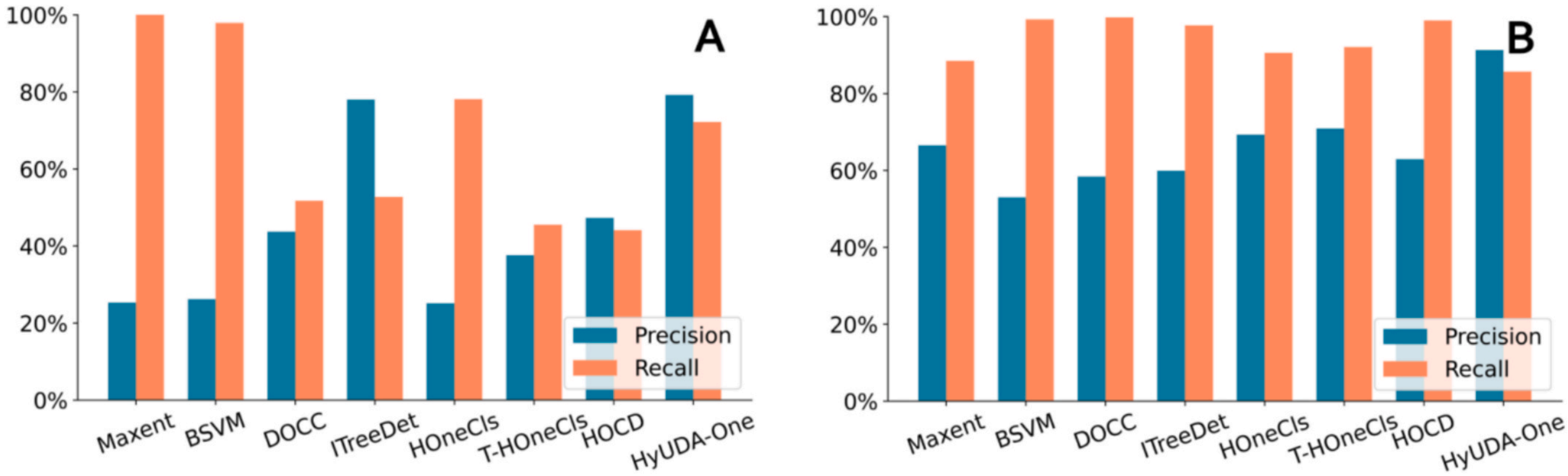

**Fig. 6.** The precision and recall (percentages) of the OCC classifiers for mapping (a) *Senegalia mellifera* and (b) *Vachellia tortillis* in Lumo.

*Senegalia mellifera*, FPGA and MLUDA obtained F1-scores of 0.402 and 0.191, respectively, compared with 0.788 achieved by the one-class baseline ITreeDet. For *Commiphora africana*, MLUDA achieved an F1-score of 0.348, markedly lower than 0.759 by ITreeDet. In the target domain Lumo, the performance gap became more evident in the cross-domain setting (Table 5). FPGA and the multiclass domain adaptation method MLUDA degraded substantially for *Senegalia mellifera*, while HyUDA-One achieved a much higher F1-score of 0.756. Although MLUDA obtained a relatively competitive result for *Vachellia tortilis*, HyUDA-One still provided higher performance.

### 4.4. Ablation analysis

The integration of RA and SSPPL demonstrated the most robust performance for mapping *Senegalia mellifera* and *Vachellia tortilis*. As presented in Table 6, the model trained on Lumo data (Train: Lumo) provided a upper bound on performance. Training exclusively within Choke without adaptation (Train: Choke) resulted in significantly decreased performance when applied to the target domain. For *Senegalia mellifera*, introducing RA improved the F1-score from 0.629 to 0.696, the AUC from 0.777 to 0.859 and the AP from 0.642 to 0.742. Applying PPL alone slightly decreased the F1-score but improved the AUC to 0.863. Combining RA and PPL further increased the AUC to 0.884. Incorporating the SSPPL notably enhanced the model performance, achieving an F1-score of 0.756, AUC of 0.890, and AP of 0.796. In addition, the balance between precision and recall gradually improved with the integration of various components (Fig. 9), verifying the effectiveness of the proposed modules. A similar tendency was observed for *Vachellia tortilis*. Using RA alone yielded only marginal improvements. In contrast, PPL effectively increased the model performance, with an F1-score of 0.854, AUC of 0.896 and AP of 0.907. The combined RA + PPL module further improved the AP to 0.935. Ultimately, adding the SSPPL produced the highest accuracy, achieving an F1-score of 0.884, AUC of 0.943, and AP of 0.958, along with superior precision and recall.

### 4.5. t-SNE analysis

The t-SNE analysis verified that the sequential integration of RA, PPL, and SSPPL modules progressively improved the discriminability of deep features. For *Senegalia mellifera* (Fig. 10a–10d, 10i), the initial feature representation without domain adaptation (Fig. 10a) showed clear overlap between the target tree species and negative classes, indicating the challenge of discriminating them under domain shift. RA (Fig. 10b) or PPL (Fig. 10c) alone led to marginal improvements in feature separation. Combining RA and PPL modules (Fig. 10d) notably increased the class separability. Ultimately, incorporating the SSPPL (Fig. 10i) resulted in the clearest distinction with minimal overlap between the positive and negative classes. Similarly, for *Vachellia tortilis* (Fig. 10e–10h, 10j), the initial scenario (Fig. 10e) exhibited overlapping feature distributions between classes. The addition of RA (Fig. 10f) slightly improved separation, while PPL (Fig. 10g) effectively enhanced class discrimination. Integrating both RA and PPL (Fig. 10h) further improved feature separation, and the introduction of the SSPPL (Fig. 10j) produced optimal separation with clear feature clusters.

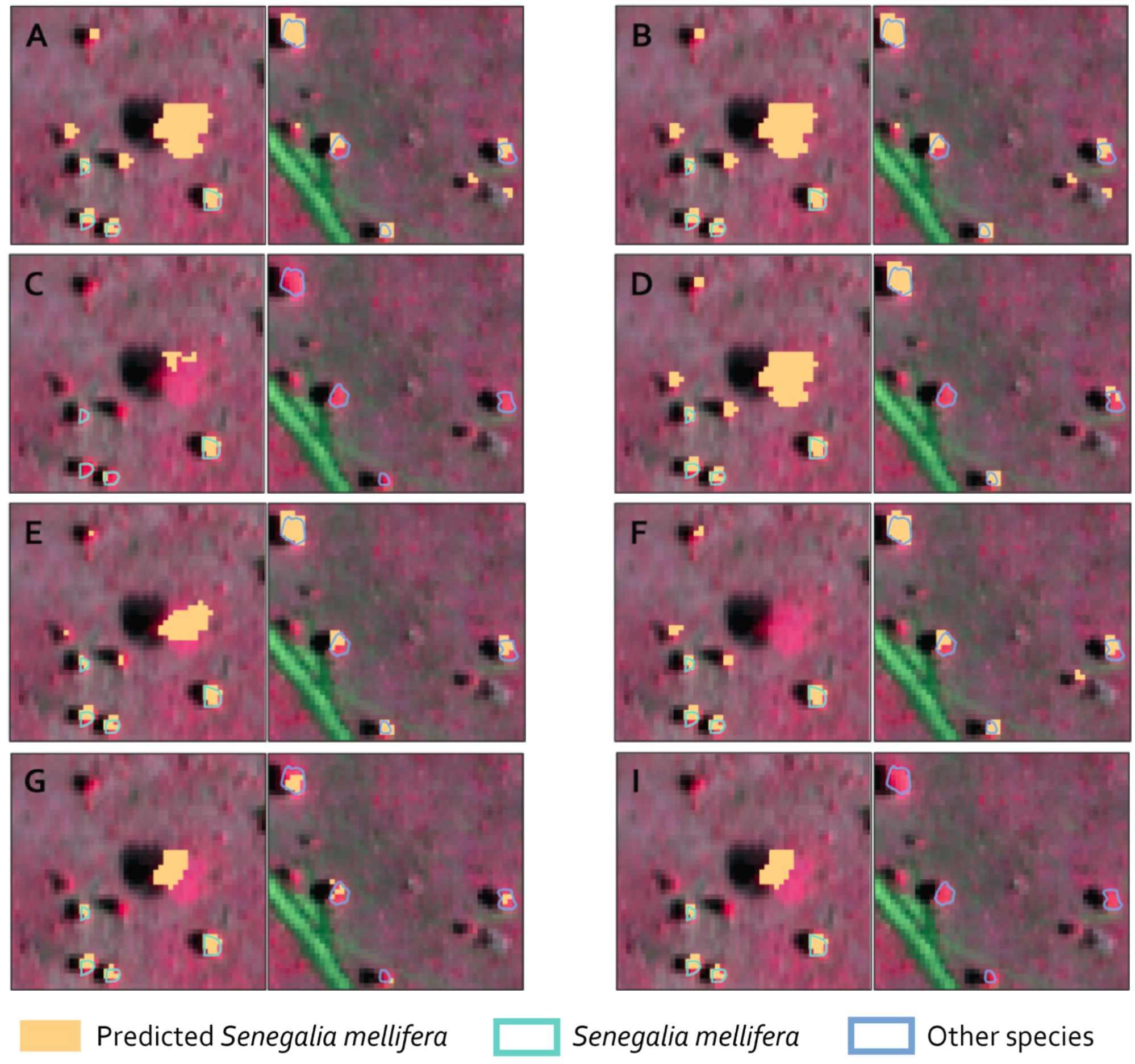

**Fig. 7.** Two local subsets of prediction maps for *Senegalia mellifera* in Lumo are shown in the left and right panels: (a) Maxent, (b) BSVM, (c) DOCC, (d) ITreeDet, (e) HOneCls, (f) T-HOneCls, (g) HOCD, and (h) HyUDA-One. The ground truth crowns of *Senegalia mellifera* are delineated as cyan polygons, the other tree species as blue polygons, and the predicted *Senegalia mellifera* by each method are shown as orange raster patches. (For interpretation of the references to colour in this figure legend, the reader is referred to the web version of this article.)

### 4.6. Melliferous tree species prediction

The prediction results revealed clear differences in the spatial distribution of the melliferous tree species between Choke and Lumo. *Senegalia mellifera* displayed a widespread and dispersed distribution pattern in Choke, with several high-density clusters concentrated in the northeastern and central areas, covering approximately 0.33 km$^2$ (Fig. 11a). *Vachellia tortilis* was relatively sparse, primarily located in the eastern part of the site, with a total coverage of about 0.06 km$^2$ (Fig. 11b). *Commiphora africana* showed a concentrated distribution, predominantly scattered across the northeastern and southeastern zones, occupying around 0.31 km$^2$ (Fig. 11c). In Lumo, *Senegalia mellifera* exhibited a significantly lower density, occupying only about 0.02 km$^2$ (Fig. 12a). *Vachellia tortilis*, on the other hand, dominated with a broad spatial distribution and substantial coverage of around 0.84 km$^2$ (Fig. 12b). Based on the CHM maps, the total tree-covered areas in Choke and Lumo were 9.33 km$^2$ and 1.88 km$^2$, respectively. The corresponding percentages of *Senegalia mellifera*, *Vachellia tortilis*, and *Commiphora africana* in Choke were 3.5 %, 0.6 % and 3.3 %, respectively. In Lumo, *Senegalia mellifera* and *Vachellia tortilis* accounted for 1.1 % and 44.7 % of the total tree cover, respectively.

### 4.7. Spatial distributions of nectar source availability

Several promising areas with abundant nectar sources were identified in both Choke and Lumo. In Choke, the nector sources for honeybees exhibited distinct spatial clusters, primarily concentrated in the northeastern area (Fig. 13a). Similarly, the nectar sources for stingless bee were also aggregated in the northeast, though with a slightly different spatial pattern (Fig. 13b). Additional high-scoring zones were observed in the southeastern part of the site. These areas overlapped with existing road networks and scattered buildings. In Lumo, nectar sources were more widely distributed, with the high-scoring areas concentrated in the northeastern zones (Fig. 14). These areas were surrounded by densely distributed buildings and accessible roads, indicating great potential for beekeeping development.

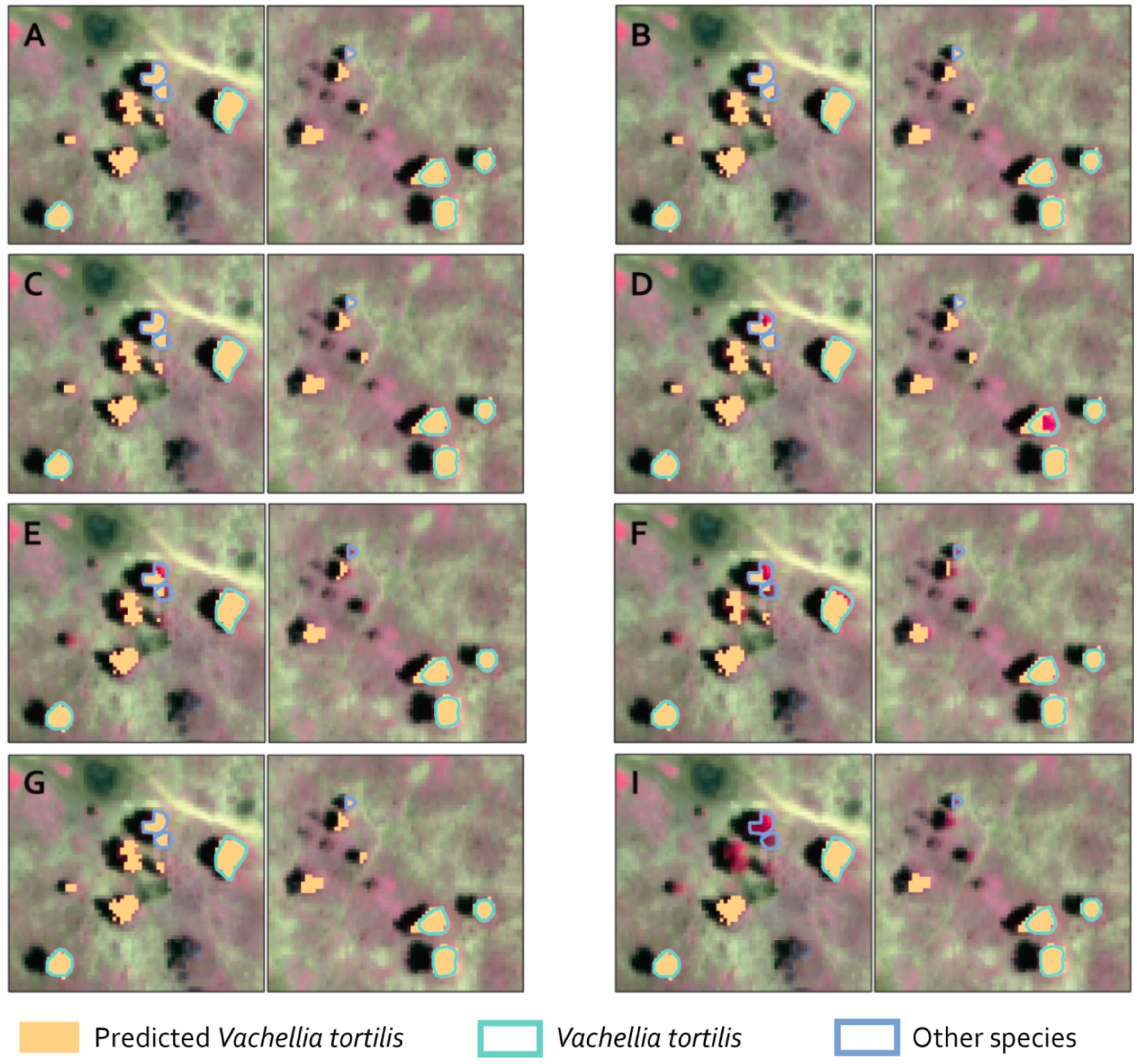

**Fig. 8.** Two local subsets of prediction maps for *Vachelia tortillis* in Lumo are shown in the left and right panels: (a) Maxent, (b) BSVM, (c) DOCC, (d) ITreeDet, (e) HOneCls, (f) T-HOneCls, (g) HOCD, and (h) HyUDA-One. The ground truth crowns of *Vachelia tortillis* are delineated as cyan polygons, the other tree species as blue polygons, and the predicted *Vachelia tortillis* by each method are shown as orange raster patches. (For interpretation of the references to colour in this figure legend, the reader is referred to the web version of this article.)

**Table 3**
Generalizability of the OCC classifiers for mapping *Senegalia mellifera* and *Vachelia tortillis* in Lumo. The best performance is marked with bold.

| Method | *Senegalia mellifera* | | | *Vachellia tortilis* | | |
|---|---|---|---|---|---|---|
| | F1-score | AUC | AP | F1-score | AUC | AP |
| Maxent | 0.404 | 0.398 | 0.197 | 0.760 | 0.735 | 0.779 |
| BSVM | 0.413 | 0.407 | 0.202 | 0.691 | 0.473 | 0.494 |
| DOCC | 0.474 | 0.736 | 0.509 | 0.737 | 0.483 | 0.524 |
| ITreeDet | 0.629 | 0.777 | 0.642 | 0.743 | 0.471 | 0.547 |
| HOneCls | 0.380 | 0.671 | 0.249 | 0.785 | 0.806 | 0.826 |
| T-HOneCls | 0.412 | 0.616 | 0.323 | 0.801 | 0.824 | 0.821 |
| HOCD | 0.457 | 0.631 | 0.399 | 0.770 | 0.553 | 0.657 |
| **HyUDA-One** | **0.756** | **0.890** | **0.796** | **0.884** | **0.943** | **0.958** |

**Table 5**
F1-score, AUC, and AP of the compared multiclass and one-class classification methods for mapping melliferous tree species in the target domain (Lumo). The best performance is marked in bold.

| Method | *Senegalia mellifera* | | | *Vachellia tortilis* | | |
|---|---|---|---|---|---|---|
| | F1-score | AUC | AP | F1-score | AUC | AP |
| FPGA | 0.361 | 0.739 | 0.295 | 0.538 | 0.628 | 0.679 |
| MLUDA | 0.107 | 0.398 | 0.043 | 0.737 | 0.459 | 0.639 |
| **HyUDA-One** | **0.756** | **0.89** | **0.796** | **0.884** | **0.943** | **0.958** |

**Table 4**
F1-score, AUC, and AP of the compared multiclass and one-class classification methods for mapping melliferous tree species in the source domain (Choke). The best performance is marked in bold.

| Method | *Senegalia mellifera* | | | *Vachellia tortilis* | | | *Commiphora africana* | | |
|---|---|---|---|---|---|---|---|---|---|
| | F1-score | AUC | AP | F1-score | AUC | AP | F1-score | AUC | AP |
| FPGA | 0.402 | 0.763 | 0.423 | 0.661 | 0.896 | 0.699 | 0.693 | 0.861 | 0.650 |
| MLUDA | 0.191 | 0.691 | 0.318 | 0.838 | 0.982 | 0.917 | 0.348 | 0.656 | 0.414 |
| ITreeDet (OCC) | **0.788** | **0.942** | **0.848** | **0.845** | **0.993** | **0.931** | **0.759** | **0.949** | **0.712** |

**Table 6**
The F1-score, AUC, and AP performance of the models with various modules for mapping *Senegalia mellifera* and *Vachelia tortillis* in Lumo. The best performance is marked with bold.

| OCC Model | *Senegalia mellifera* | | | *Vachellia tortilis* | | |
|---|---|---|---|---|---|---|
| | F1-score | AUC | AP | F1-score | AUC | AP |
| Train: Lumo | 0.871 | 0.99 | 0.95 | 0.918 | 0.962 | 0.967 |
| Train: Choke | 0.629 | 0.777 | 0.642 | 0.743 | 0.471 | 0.547 |
| RA | 0.696 | 0.859 | 0.742 | 0.722 | 0.446 | 0.552 |
| PPL | 0.617 | 0.863 | 0.682 | 0.854 | 0.896 | 0.907 |
| RA + PPL | 0.657 | 0.884 | 0.716 | 0.844 | 0.91 | 0.935 |
| **RA + SSPPL** | **0.756** | **0.89** | **0.796** | **0.884** | **0.943** | **0.958** |

## 5. Discussion

### 5.1. The performance of the one-class classification methods

Several existing OCC methods were evaluated in this study for mapping nectar-producing tree species using airborne hyperspectral imagery and LiDAR data in Kenya. The results in Choke demonstrated considerable variations in performance across the methods. Among traditional OCC algorithms, the P classifier Maxent achieved comparable accuracy to the PU classifier BSVM, despite not utilizing any unlabeled samples. Among deep learning-based PU methods, the patch-free methods HOneCls and T-HOneCls exhibited high classification accuracy for mapping *Senegalia mellifera* and *Commiphora africana*, but slightly underperformed for *Vachellia tortilis* compared to other approaches. In contrast, the patch-based methods DOCC and ITreeDet consistently

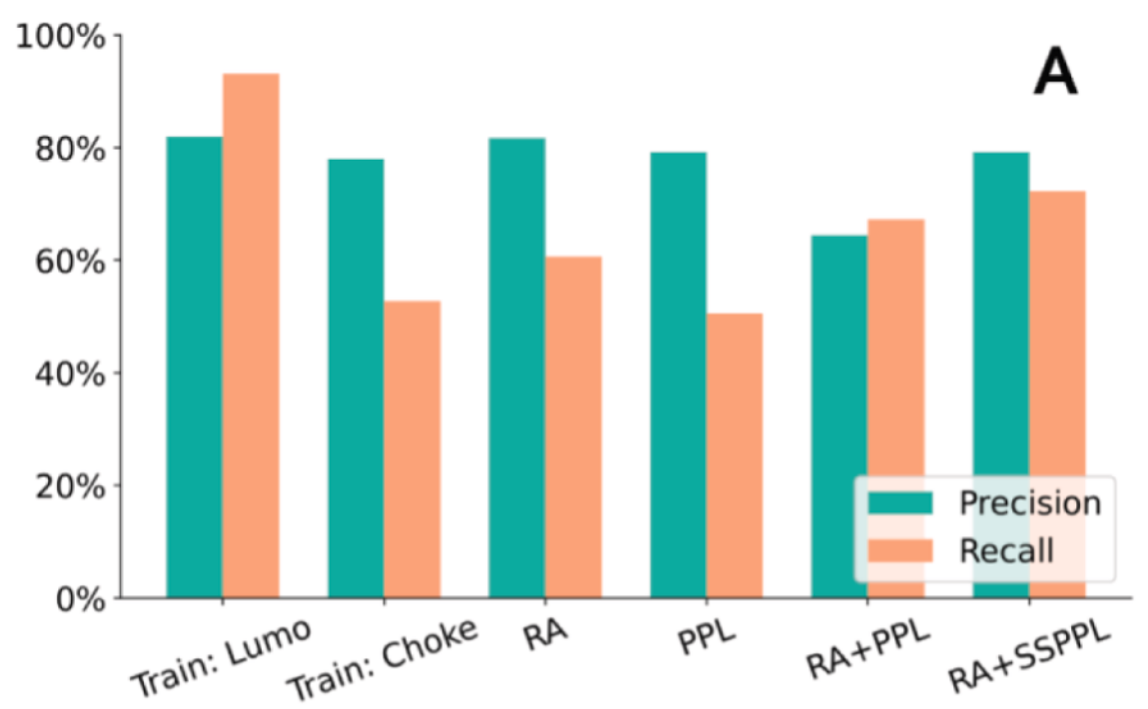

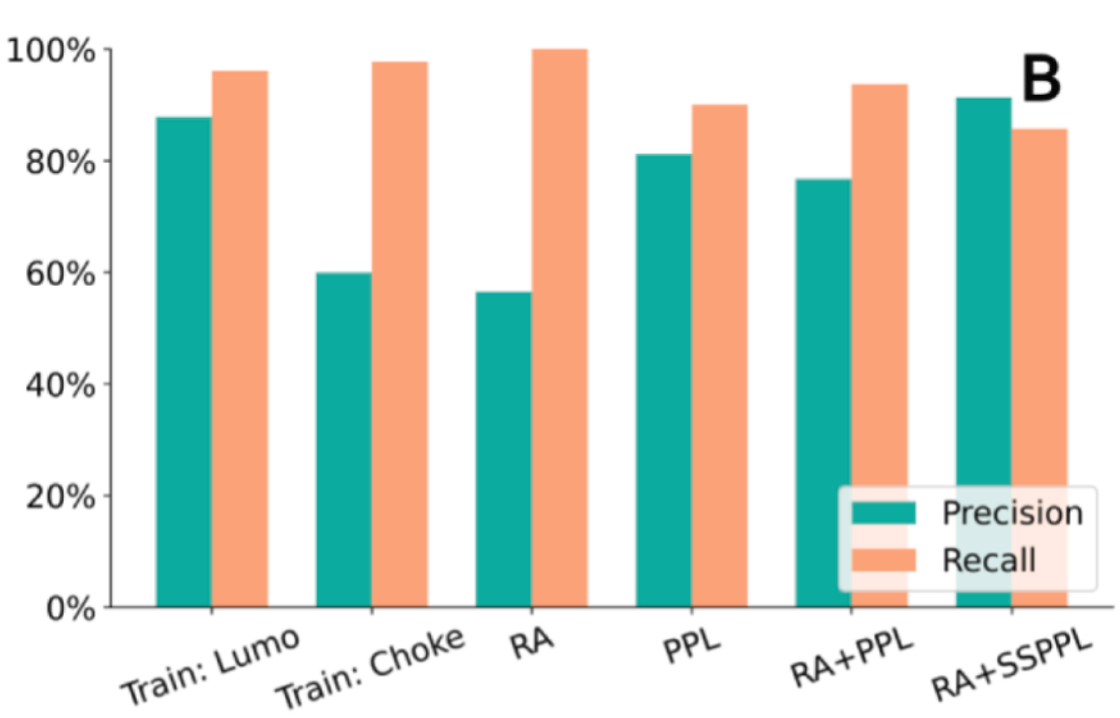

**Fig. 9.** The precision and recall (percentages) of the models with various modules for mapping (a) *Senegalia mellifera* and (b) *Vachelia tortillis* in Lumo.

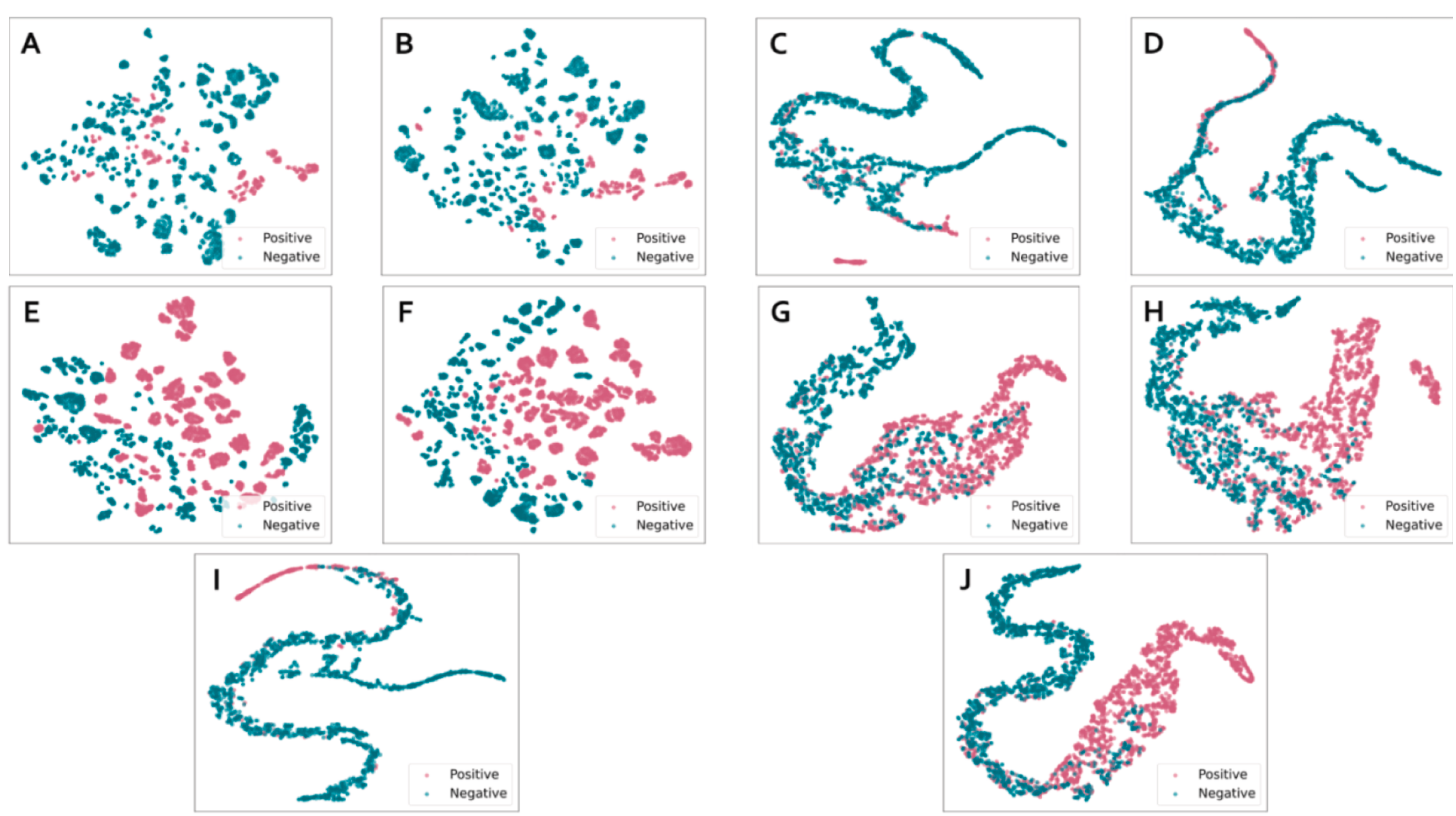

**Fig. 10.** T-SNE visualization of deep features extracted from the final hidden layer of the networks with various modules for mapping *Senegalia mellifera* and *Vachellia tortilis* in Lumo. (a) and (e) show the visualization result of the model without any processing. (b)-(d) and (i) are visualization results of RA, PPL, RA + PPL, and RA + SSPPL models for *Senegalia mellifera*, respectively. Correspondingly, (f)-(h) and (j) visualize results of various models for *Vachellia tortilis*.

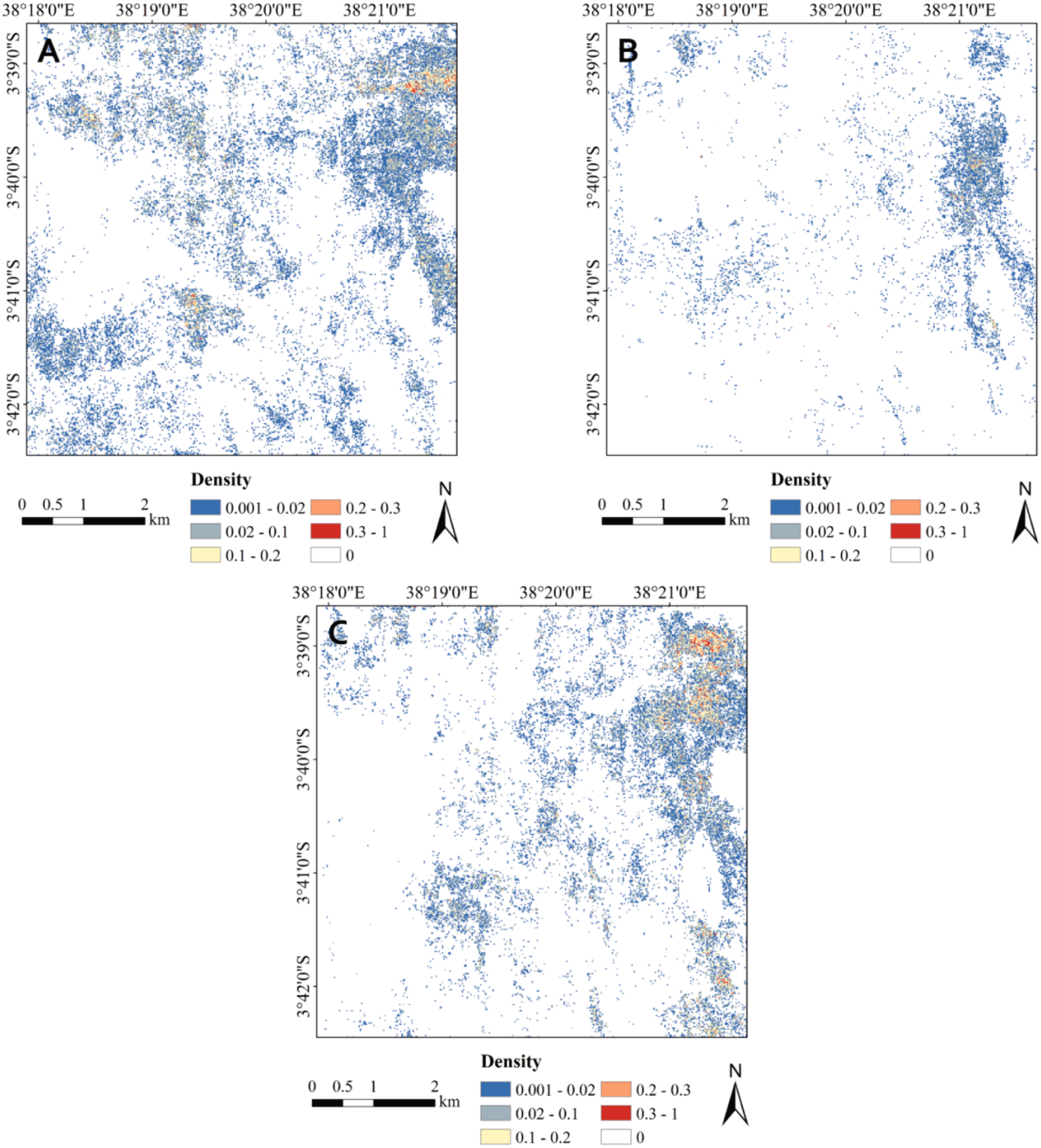

**Fig. 11.** Tree species cover of the best model for (a) *Senegalia mellifera*, (b) *Vachellia tortilis*, and (c) *Commiphora africana* in Choke in the 20-m grid cells.

achieved the best overall accuracy across all three species.

Previous studies have provided valuable references for assessing model performance in mapping tasks. Comparable performance between Maxent and BSVM was reported in mapping *Centaurea solstitalis* and *Phalaris aquatica* (Skowronek et al. 2017). Maxent achieved precision/recall values of 0.73/0.89 and 0.71/0.50, while BSVM yielded values of 0.76/0.76 and 0.52/0.53, respectively. Additionally, BSVM achieved relatively higher F1-scores of 0.76 and 0.78 for mapping invasive tree species *Eucalyptus* spp. and *Acacia mearnsii* (Piiroinen et al. 2018). Higher classification accuracy compared to this study was reported for DOCC (Lei et al. 2021), where DOCC was applied for crop classification using satellite hyperspectral imagery (Zhuhai-1), achieving F1-scores of 0.874 for winter wheat and 0.877 for rapeseed. These results suggest that crop classification may benefit from greater spectral distinctiveness and less variability compared to tree species mapping (Roth et al. 2015). The performance of ITreeDet in this study aligns closely with the results reported in the previous study (Zhao et al. 2022), where F1-scores of 0.86 for *Eucalyptus* spp. and 0.70 for *Acacia mearnsii* were achieved, further validating the robustness of this method for tree species mapping. In contrast, significantly higher accuracy of HOneCls and T-HOneCls, with F1-scores above 0.936 and 0.945 were obtained, respectively, (Zhao et al. 2023a; Zhao et al. 2023b) in multi-class crop mapping tasks. This further highlights spectral overlap, spatial heterogeneity, and high intra-class variability pose significant challenges for tree species mapping (Fassnacht et al. 2016).

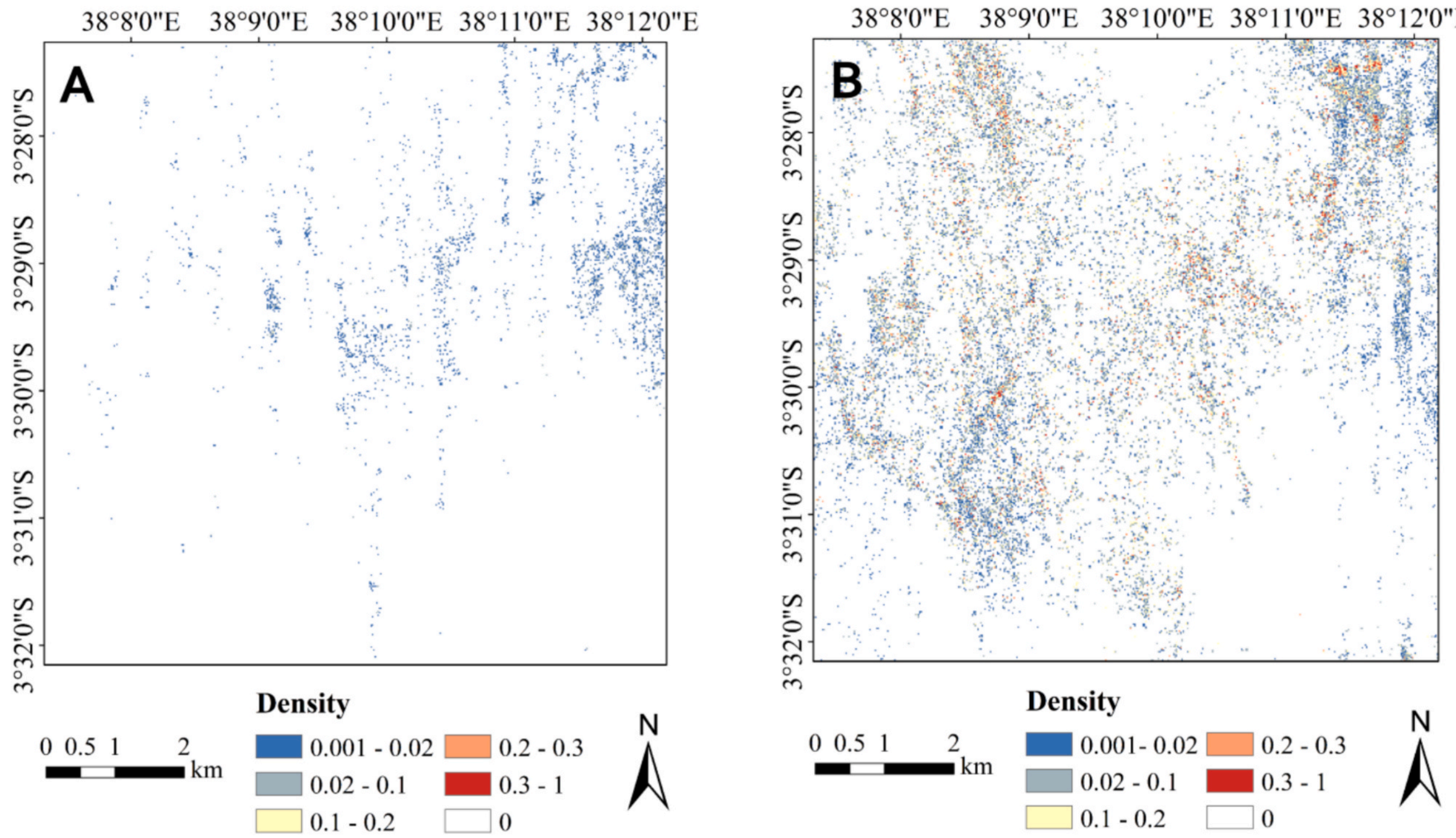

**Fig. 12.** Tree species cover of the best model for (a) *Senegalia mellifera* and (b) *Vachellia tortilis* in Lumo in the 20-m grid cells.

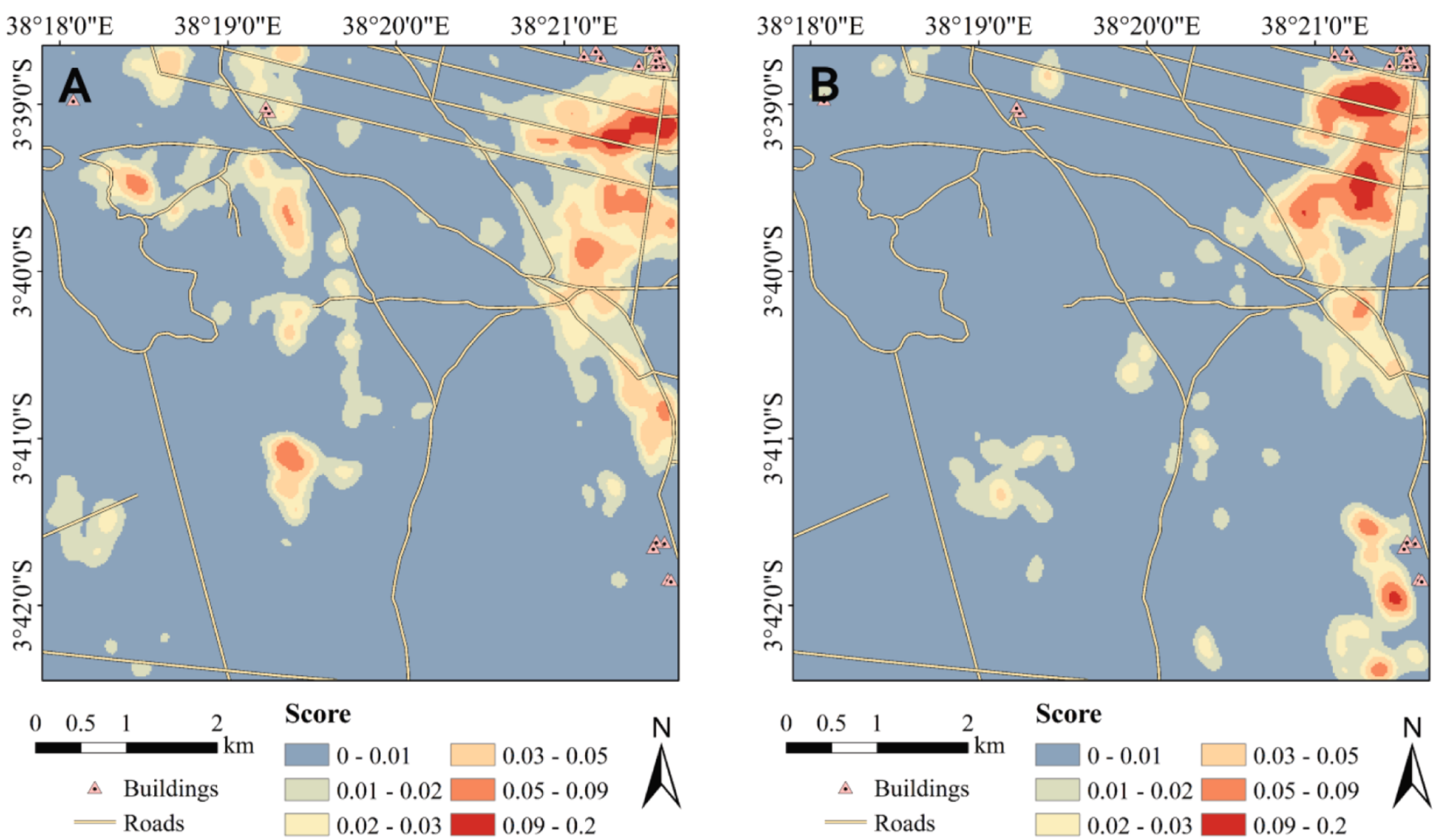

**Fig. 13.** Spatial distributions of nectar source availability for (a) honeybees and (b) stingless bees in Choke. Building locations and road infrastructure are overlaid to indicate suitability for potential apiary development. Road data sourced from the Humanitarian OpenStreetMap Team.

### *5.2. The generalization capability of the one-class classification methods*

The generalization capability of the existing OCC methods and the proposed HyUDA-One framework was explored by evaluating their performance in Lumo. It is worth noting that direct training and testing within Lumo (Train: Lumo in Table 3) achieved considerably higher accuracy than the source-domain experiment where both training and testing were conducted within Choke (ITreeDet in Table 2), suggesting that tree species classification in Lumo was less challenging. The spatial distribution of trees in Lumo was generally dispersed, while more continuous tree cover was observed in Choke. Such differences in spatial distribution patterns may contribute to variations in mapping complexity and model performance across the two sites. In addition to these structural and ecological differences, domain shift can also be exacerbated by variations in hyperspectral survey parameters, such as flight height, illumination geometry and atmospheric conditions, which can introduce systematic spectral discrepancies between the two campaigns even after standard preprocessing.

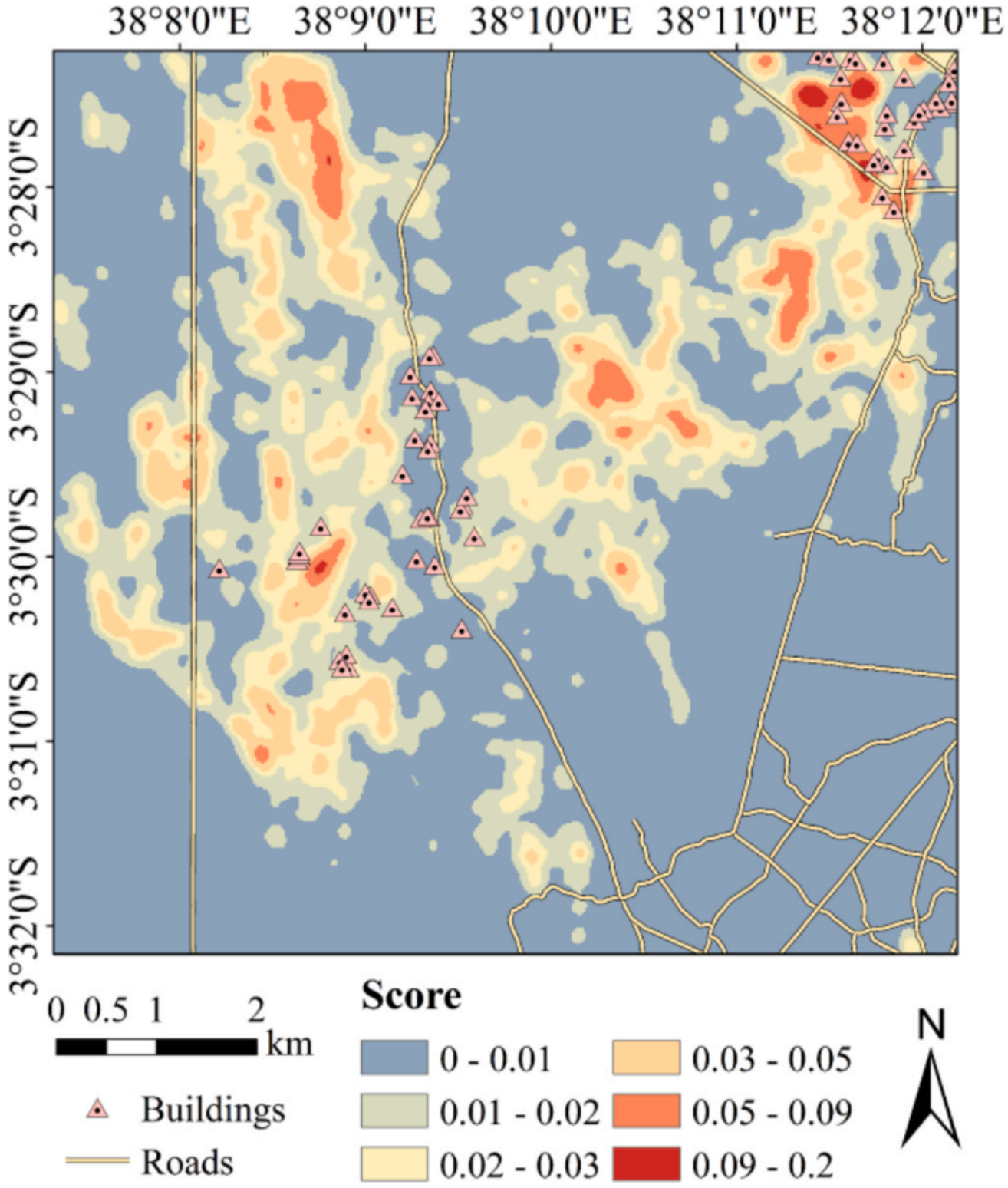

**Fig. 14.** Spatial distribution of honeybees nectar sources in Lumo. Buildings and road networks are overlaid for reference.

The results of the two species in Lumo exhibited different performance trends. For *Senegalia mellifera*, the accuracy of traditional OCC methods Maxent and BSVM severely decreased, indicating high susceptibility to domain shift. All deep OCC classifiers also showed varying degrees of performance degradation. In contrast, the proposed HyUDA-One framework significantly improved the mapping accuracy, achieving an F1-score of 0.756, demonstrating its effectiveness in mitigating domain shift challenges. For *Vachellia tortilis*, Maxent, and the patch-free methods HOneCls and T-HOneCls exhibited relatively stable generalization capability, while BSVM, and the patch-based approaches DOCC and ITreeDet displayed considerable substantial performance declines. HyUDA-One markedly enhanced the mapping performance in Lumo. Overall, the traditional PU method BSVM and the patch-based deep OCC methods exhibited poorer generalization capability across geographically distinct domains.

While there is large literature on applying OCC methods to hyperspectral classification of crops and tree species, few studies have focused on unsupervised domain adaptation in this field, which is crucial for practical large-scale mapping tasks. Prior research has highlighted the necessity of transfer learning in OCC when the data availability in target domains is limited (Chen and Liu 2014). When there are positive labels in target domains, using OCSVM in target domains can significantly reduce the demand for collection of negative samples (Chen and Liu 2014). Similarly, a kernel-adaptation-based OCSVM (Xue and Beauseroy 2017) has been developed to adjust pre-trained parameters during transfer learning. Adversarial learning has also been employed to generate pseudo-negative samples (Ding et al. 2023) and to align feature distributions between source and target domains (Chi and Mao 2024; Mao et al. 2023). However, these previous frameworks are primarily designed for scenarios where positive samples from target domains are available. In such cases, OCC models can achieve good performance in tree species mapping, as confirmed in our experiments. The complete absence of labeled samples in target domains presents a more challenging scenario. This study provides an effective strategy to improve the generalization capability of hyperspectral OCC classifiers under such label-scarce conditions. This fills a significant gap in the existing literature and provides practical value for tree species ecological mapping and management.

The various domain adaptation modules were evaluated through the ablation analysis. RA was implemented by adjusting the spectral reflectance of Lumo based on the mean and variance of Choke. This approach led to noticeable improvements for mapping *Senegalia mellifera*, indicating that part of the domain shift may be attributed to systematic spectral differences between the two sites (Mohammadi et al. 2024). However, RA had limited impact on mapping accuracy for *Vachellia tortilis*. This result suggests that the domain shift for this species arises from more complex factors beyond mere spectral differences, possibly related to environmental variations affecting growth conditions between landscapes (Everest et al. 2025). Subsequently, applying PPL alone within Lumo yielded minor accuracy improvements for *Senegalia mellifera*, while it produced remarkable accuracy gains for *Vachellia tortilis*. It is likely that the PPL strategy successfully identified a representative subset of *Vachellia tortilis* samples. Iterative adaptation based on these correct pseudo labels subsequently improved the model generalizability significantly. Combining RA with PPL further improved classification accuracy for both species. Nonetheless, potential errors accumulation in the pseudo labels still limited the model performance. Ultimately, introducing the SSPPL significantly improved model accuracy, clearly demonstrating the effectiveness of exploiting non-local spatial–spectral contextual information. Erroneous pseudo labels were effectively suppressed by enforcing spatial consistency and spectral similarity constraints, further optimizing the model predictions.

### 5.3. The mapping on melliferous tree species

Recent studies have demonstrated the feasibility of mapping the cover of *Senegalia mellifera* using multiclass classification, while species-level mapping of *Vachellia tortilis* and *Commiphora africana* remains to be explored. On the one hand, EnMAP hyperspectral satellite data (30 m resolution) and multitemporal Sentinel-2 imagery have been combined to map *Senegalia mellifera* in a South African savanna (Karakizi et al. 2024). Similarly, Harkort et al. (2025) introduced a novel method using Sentinel-2 time-series to exploit seasonal phenological differences and map fractional woody cover, including *Senegalia mellifera*, in Sub-Saharan rangelands. These studies relied on labels of multiple tree species for model training, while only labeled *Senegalia mellifera* samples were required in this study. On the other hand, previous research tended to group *Vachellia* spp. into a single category for classification (Karakizi et al. 2024). In addition, *Commiphora* spp. often co-occurs with *Acacias* spp.. They are usually categorized as part of *Acacia-Commiphora* woodland complexes rather than as separate species (Coe 1978). For instance, broad-scale vegetation surveys in the Tsavo region of Kenya historically identified *Acacia-Commiphora* bushland as a dominant vegetation type (Kamau 2017). This study successfully mapped the three melliferous tree species individually using airborne hyperspectral imagery and LiDAR data. Superior performance was achieved through the proposed HyUDA-One framework without relying on extensive data from other classes. A recent study reported that *Vachellia tortilis* contributes 64 % of the aboveground biomass in the grassland system of Taita Hills (Amara et al. 2023). This aligns well with the mapping result in Lumo, where *Vachellia tortilis* contributed 44.7 % of the total tree cover. Despite these encouraging results, several challenges remain to be addressed. The reflectance spectra of the target species can partly overlap with those of other broad-leaved trees, especially in drier periods when crowns are partially leafless or foliage is senescent, which reduces species-specific spectral contrast. These characteristics indicate the potential benefits of considering phenology and multi-temporal imagery for tree species mapping applications. In addition, few *Commiphora africana* samples were available in Lumo, suggesting that further targeted field data collection could be useful for validating its generalization capability.

Beyond Africa, many important nectar-producing tree species have been the subject of remote sensing mapping, often driven by forestry or invasive species management interests. UAV-based high-resolution imagery has been utilized to map flowering *Robinia pseudoacacia* in Europe (Atanasov et al. 2024) and *Prosopis glandulosa* in North American rangelands (Jackson et al., 2020). Meanwhile, IKONOS multispectral satellite imagery achieved approximately 86 % accuracy in mapping *Melaleuca quinquenervia* in Florida's Everglades (Fuller, 2005). The mapping accuracy achieved in this study using airborne hyperspectral imagery is comparable to the results reported from these diverse platforms. A slight decrease in accuracy is expected when using satellite data at 10–30 m scale, as satellite-scale pixels may mix signals from multiple objects. UAV-based hyperspectral imagery offers a promising alternative for small-scale precise tree species detection, due to its lower cost, high flexibility, and higher spatial resolution. Additionally, all three target species had foliage during the hyperspectral data acquisition in this study. In the future, it may be considered to collect hyperspectral data during the flowering period of the three tree species to obtain more discriminative features.

The detailed mapping of three nectar-producing tree species provides practical information of the nectar source availability, directly supporting the strategic planning for local beekeeping development. Specifically, Choke represents a multi-functional landscape encompassing wildlife conservation, livestock grazing, carbon sequestration initiatives, and sustainable mining (Abera et al. 2022). Although our results demonstrated considerable potential for beekeeping in Choke, the site remains largely uninhabited. Consequently, significant initial investment and sustained maintenance would be necessary to develop and manage a productive apicultural system within Choke. Lumo is characterized by both livestock herding and wildlife conservation (Sorokina et al. 2024). The results of nectar source availability indicate notable beekeeping potential in Lumo. However, ongoing overgrazing poses a threat to ecological stability in this landscape, contributing to environmental degradation and biodiversity loss (Sorokina et al. 2024). Beekeeping has been recognized in recent studies for their role in fostering ecological restoration, mitigating degradation, and promoting biodiversity conservation in pastoral landscapes (Tonietto and Larkin 2018). Therefore, expanding apiaries based on the nectar source assessments in this sudy could effectively support ecological restoration while simultaneously generating economic benefits for local communities.

## 6. Conclusions

In this study, a hyperspectral OCC framework based on unsupervised domain adaptation was proposed for tree species mapping. The spatial–spectral regularized pseudo-positive learning was designed to address domain shift of geographically distinct regions. To the best of our knowledge, HyUDA-One represents one of the first explorations of unsupervised domain adaptation–based one-class classification frameworks for hyperspectral tree species mapping. The proposed HyUDA-One framework can effectively improve generalization capability without the need for extra labeled samples when mapping tree species in new domains. The effectiveness of HyUDA-One was demonstrated for mapping three key melliferous tree species, i.e., *Senegalia mellifera*, *Vachellia tortilis*, and *Commiphora africana*, in two savanna landscapes in Kenya. Differences in the mapping results between landscapes may reflect variations in tree canopy structure, which could be influenced by differing levels of herbivore pressure. Furthermore, the analysis of nectar source availablity for honeybees and stingless bees based on the densities of melliferous tree species provides a data-driven reference for local beekeeping management. The proposed HyUDA-One framework can be also utilized for other remote sensing mapping tasks, e.g., invasive species detection, to increase mapping generalization capability to new domains.

## Declaration of Generative AI and AI-assisted technologies in the writing process

During the preparation of this work the authors used ChatGPT 4o in order to improve the readability and language of the manuscript. After using this tool, the authors reviewed and edited the content as needed and take full responsibility for the content of the publication.

## CRediT authorship contribution statement

**Zhaozhi Luo:** Writing – original draft, Visualization, Validation, Software, Resources, Methodology, Investigation, Formal analysis, Data curation, Conceptualization. **Janne Heiskanen:** Writing – review & editing, Supervision, Resources, Methodology, Conceptualization. **Ilja Vuorinne:** Writing – review & editing, Software, Resources, Methodology. **Ian Ocholla:** Writing – review & editing, Resources. **Shiqi Zhang:** Writing – review & editing, Formal analysis. **Saana Järvinen:** Writing – review & editing, Investigation. **Xinyu Wang:** Writing – review & editing, Supervision. **Yanfei Zhong:** Writing – review & editing, Supervision. **Petri Pellikka:** Writing – review & editing, Supervision, Project administration, Funding acquisition, Conceptualization.

## Declaration of competing interest

The authors declare that they have no known competing financial interests or personal relationships that could have appeared to influence the work reported in this paper.

## Acknowledgements

This work was funded by the European Union DG International Partnerships under the DeSIRA program (FOOD/2020/418-132) through the ESSA project, PI Petri Pellikka. This work was also supported by a doctoral researcher position in Atmospheric Sciences funded by the University of Helsinki and by the EDUFI Fellowship funded by the Finnish National Agency for Education. The authors would like to acknowledge the Taita Research Station of the University of Helsinki for providing logistics and accommodation during fieldwork. We are grateful to Mwadime Mjomba, Peter Mwasi, and Darius Kimuzi for their invaluable assistance in data collection. We acknowledge research licenses from the National Commission for Science, Technology and Innovation (NACOSTI/P/21/14537, NACOSTI/P/21/14977, and NACOSTI/P/25/4172935).